\documentclass[a4paper,11pt]{article}
\pdfoutput=1
\usepackage{amssymb,amsmath,amsfonts,makeidx,placeins,pbox,multirow}
\usepackage{graphicx,rotate,subcaption,color,slashed,cite,caption,epstopdf,verbatim}
\usepackage{longtable,tabu}
\usepackage{array}
\usepackage{color}
\usepackage{url}
\usepackage[table]{xcolor}
\usepackage{placeins}
\usepackage{amsmath}
\usepackage{mathtools}
\usepackage{amssymb}
\usepackage[colorlinks=true,
            linkcolor=blue,
            citecolor=blue,
            urlcolor=blue]{hyperref}
\usepackage[normalem]{ulem}
\usepackage{graphicx}
\usepackage{svg}
\usepackage[justification=centering]{caption}
\usepackage{subcaption}
\usepackage{booktabs}
\usepackage{makecell}  
\usepackage{arydshln}
\newcolumntype{P}[1]{>{\centering\arraybackslash}p{#1}}
\usepackage{tablefootnote}

\def\lapp{\mathrel{\rlap{\raise.5ex\hbox{$<$}}
                    {\lower.5ex\hbox{$\sim$}}}}
\def\gapp{\mathrel{\rlap{\raise.5ex\hbox{$>$}}
                  {\lower.5ex\hbox{$\sim$}}}}

\usepackage{color}
\long\def\/*#1*/{}
\usepackage{color}
 \usepackage[normalem]{ulem}
\definecolor{darkgreen}{cmyk}{1,0,1,0.4}
\definecolor{darkred}{cmyk}{0,1,1,0.4}
\definecolor{rosso}{cmyk}{0,1,1,0.4}
\definecolor{rossos}{cmyk}{0,1,1,0.55}
\definecolor{rossoc}{cmyk}{0,1,1,0.2}
\definecolor{blu}{cmyk}{1,1,0,0.3}
\definecolor{blus}{cmyk}{1,1,0,0.6}
\definecolor{bluc}{cmyk}{1,1,0,0.1}
\definecolor{verde}{cmyk}{0.92,0,0.59,0.25}
\definecolor{verdec}{cmyk}{0.92,0,0.59,0.15}
\definecolor{verdes}{cmyk}{0.92,0,0.59,0.4}
\definecolor{grigio}{cmyk}{0,0,0,0.07}
\definecolor{rosa}{cmyk}{0,0.1,0.1,0.02}
\definecolor{rosino}{cmyk}{0,0.05,0.05,0.02}
\definecolor{rosas}{cmyk}{0,0.3,0.25,0.05}
\definecolor{celeste}{cmyk}{0.1,0,0,0.02}
\definecolor{giallino}{cmyk}{0,0,0.4,0.02}
\definecolor{rosso}{cmyk}{0,1,1,0.4}
\definecolor{rossos}{cmyk}{0,1,1,0.55}
\definecolor{rossoc}{cmyk}{0,1,1,0.2}
\definecolor{blu}{cmyk}{1,1,0,0.3}
\definecolor{bluc}{cmyk}{1,1,0,0.1}
\definecolor{blucc}{cmyk}{0.7,0.5,0,0}
\definecolor{viola}{cmyk}{0,1,0,0.6}
\definecolor{viola2}{cmyk}{0,1,0.2,0.6}
\definecolor{verde}{cmyk}{0.92,0,0.59,0.25}
\definecolor{verdec}{cmyk}{0.92,0,0.59,0.15}
\definecolor{verdes}{cmyk}{0.92,0,0.59,0.4}
\definecolor{verdino}{cmyk}{0.12,0,0.09,0.05}
\definecolor{giallo}{cmyk}{0,0,1,0}
\definecolor{gialloverde}{cmyk}{0.44,0,0.74,0}

\evensidemargin=\oddsidemargin
\def\bar {\overline}

\def\bea {\begin{eqnarray}}
\def\eea {\end{eqnarray}}

\def\beq{\begin{equation}}
\def\eeq{\end{equation}}
\def\barr{\begin{array}}
\def\earr{\end{array}}

\def\beq{\begin{equation}}
\def\eeq{\end{equation}}

\newcommand{\nc}{\newcommand}

\nc{\hi}{H}
\nc{\hit}{\widetilde{H}}
\nc{\hij }{\mbox{${\hi^\dag i\,\raisebox{2mm}{\boldmath ${}^\leftrightarrow$}\hspace{-4mm} D_\mu\,\hi}$}}
\nc{\hijt}{\mbox{${\hi^\dag i\,\raisebox{2mm}{\boldmath ${}^\leftrightarrow$}\hspace{-4mm} D_\mu^{\,a}\,\hi}$}}

\def\gev{\rm GeV}
\def\tev{\rm TeV}

\def\gev{\,\ensuremath{\mathrm{Ge\kern -0.1em V}}}
\def\tev{\,\ensuremath{\mathrm{Te\kern -0.1em V}}}

\DeclareUnicodeCharacter{2212}{-}
\UseRawInputEncoding

\begin{document}

\begin{center}
{\Large {\bf RG evolution and effect of intermediate new-physics on $\Delta B=1$ four-fermion operators}} \\
\vspace*{0.8cm} {\sf Mathew Thomas Arun \footnote{mathewthomas@iisertvm.ac.in}, Shyam M\footnote{shyam.m.work.27@gmail.com}, Ritik Pal\footnote{ritik24@iisertvm.ac.in }} \\
\vspace{10pt} {\small } {\em  School of Physics, Indian Institute of Science Education and Research, Thiruvananthapuram 695551, Kerala, India}
\normalsize
\end{center}
\bigskip
\begin{abstract}
Motivated by the stringent experimental bounds on proton lifetime and the need for precise low-energy predictions, there has been renewed interest in the renormalization group (RG) evolution of Wilson coefficients for baryon number violating (BNV) operators and their characteristic new-physics scales. In this work, we analyze the RG running of dimension-6 four-fermion operators in the $\overline{\text{MS}}$ scheme that mediate nucleon decay channels such as $p \to e^+ \pi^0$, while systematically accounting for the impact of baryon number conserving (BNC) new-physics that can enter the theory at an intermediate scale as higher-dimensional effective field theory operator. These BNC operators mix with BNV ones at 1-loop and alter the RG flow. The running is performed from the electroweak scale up to representative intermediate scales of $10^4~\text{GeV}$, $10^6~\text{GeV}$, and $10^9~\text{GeV}$, corresponding to possible thresholds for new BNC degrees of freedom. Comparing the RG evolved coefficients with current experimental bounds on nucleon decay lifetimes, we find that the inclusion of BNC-BNV mixing, dominated by top quark loops, can significantly lower the effective proton decay scale to $\sim 10^7$ GeV, thus mitigating the need of a large desert. A Python package\footnote{Available at https://github.com/rp-winter/Nucleon-Decay-SMEFT} is provided to facilitate the RG evolution of nucleon-decay Wilson coefficients, allowing for the inclusion of generic BNC effects.

\end{abstract}

\section{Introduction}
The observed matter-antimatter asymmetry of the universe remains one of the deepest unresolved questions in fundamental physics. Along with the other Sakharov conditions, namely C and CP violation, and departure from thermal equilibrium, baryon number violating processes are essential ingredients for generating the observed imbalance. Experimental searches for such processes, particularly nucleon decay, have set extremely stringent lower limits on the proton lifetime, currently exceeding $10^{34}$~years~\cite{Super-Kamiokande:2018apg}. This result strongly constrains simple new-physics scenarios that predict baryon number violation at TeV scale. Nevertheless, the search for nucleon decay remains a cornerstone of efforts to uncover physics beyond the Standard Model (BSM). 

While baryon and lepton numbers appear as accidental global symmetries of the Standard Model (SM) at the classical level, they are explicitly broken by nonperturbative quantum effects~\cite{tHooft:1976rip}, leaving $U(1)_{B-L}$ as the only conserved combination. Though there is no \emph{a priori} reason to expect these symmetries to remain exact in BSM frameworks, many theoretical constructions, like Grand unified theories (GUTs) and related gauge extensions~\cite{Mohapatra:1980qe, Mohapatra:1980de, Mohapatra:1986dg, Mohapatra:1996pu, Babu:2008rq, Arnold:2012sd, Berezhiani:2015afa, Dev:2015uca, Allahverdi:2017edd, Fridell:2021gag, Gopalakrishna:2024qxk}, naturally accommodate them. The relevant operators contributing to proton decay typically arise at dimension-6, and their coefficients are tightly constrained by the experimental limits on nucleon lifetimes (Table.~\ref{tab:decay_channels}), pushing the characteristic mass scale of these mediators close to $10^{15}$-$10^{16}$~GeV. The next generation of large-volume detectors, such as Hyper-Kamiokande~\cite{Hyper-Kamiokande:2018ofw}, DUNE~\cite{2817610}, and JUNO, are poised to extend the sensitivity of proton lifetimes to order $10^{35}$~years, offering unprecedented discovery prospects.
\begin{table}[h]
\centering
\begin{tabular}{cccc}
\hline
\hline
Process & $\Delta(B - L)$ & $\tau$ ($10^{33}$ years) \\
\hline
\hline
$p \rightarrow \pi^0\ell^+$ & 0 & 24 [16] & ~\cite{Super-Kamiokande:2020bov} \\
$p \rightarrow \eta^0\ell^+$ & 0 & 10 [4.7] & ~\cite{Super-Kamiokande:2017gev}  \\
$p \rightarrow K^+\nu$ & 0, 2 & 6.61 & ~\cite{Mine:2016mxy} \\
$n \rightarrow \pi^-\ell^+$ & 0 & 5.3 [3.5] & ~\cite{Super-Kamiokande:2017gev}  \\
$n \rightarrow \pi^0\nu$ & 0, 2 & 1.1 & ~\cite{Super-Kamiokande:2013rwg}  \\
$p \rightarrow K^0\ell^+$ & 0 & 1 [1.6] & ~\cite{Super-Kamiokande:2012zik,Super-Kamiokande:2005lev} \\
$p \rightarrow \pi^+\nu$ & 0, 2 & 0.39 & ~\cite{Super-Kamiokande:2013rwg}  \\
$n \rightarrow \eta^0\nu$ & 0, 2 & 0.158 & ~\cite{PhysRevD.59.052004}  \\
$n \rightarrow K^0\nu$ & 0, 2 & 0.13 & ~\cite{Super-Kamiokande:2013rwg} \\
$n \rightarrow K^+\ell^-$ & 2 & 0.032 [0.057] & ~\cite{Frejus:1991ben}\\
\hline
\hline
\end{tabular}
\caption{Allowed two-body nucleon decays with the lifetime $\tau$ for both the $\Delta(B - L)$ = 0, 2 nucleon to pseudoscalar meson decay channels.}
\label{tab:decay_channels}
\end{table}

\begin{figure}[h!]
    \centering
    \includegraphics[scale=0.15]{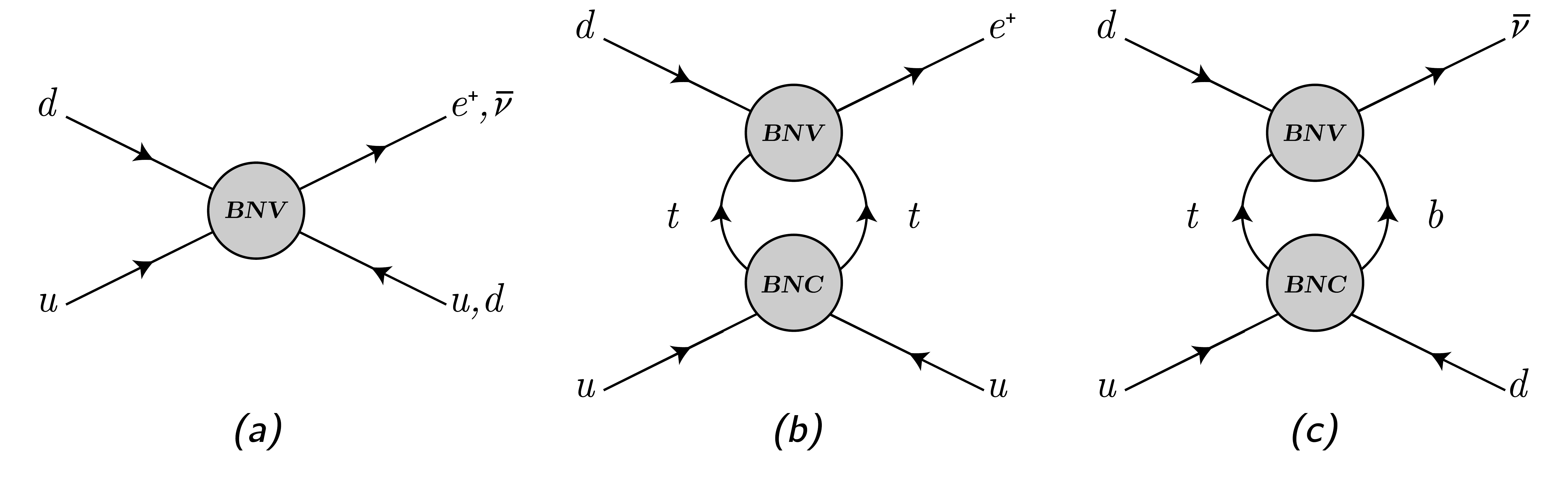}
    \caption{Representative 1-loop topologies for nucleon decay: the $(a)$ involves only BNV running, the $(b)$ loop gives the dominant (top-top) contribution to proton decay, and the $(c)$ loop gives the dominant (top-bottom) contribution to neutron decay.}
    \label{fig:block_diagram_rep}
\end{figure}

While current experimental limits offer rough estimates of the energy scale associated with baryon number violation (BNV), these estimates are incomplete because of the significant gap between the low-energy scale of measurements and the GUT scale where new-physics matches onto the BNV operators. Ongoing efforts to incorporate next-to-leading order corrections, by including 2-loop renormalisation group (RG) evolution~\cite{Hisano:2013ege,Pokorski:2017ueo,Pokorski:2019ete,Naterop:2025lzc,Banik:2025wpi} and dimension-7 operators~\cite{Beneito:2023xbk}, are essential for translating experimental searches into better constraints on the landscape of viable BNV in BSM theories. Moreover, one must also account for loop-level mixing of possible intermediate scale baryon number conserving (BNC) new-physics. Consequently, if new BNC states exist in the intermediate scale range, that interact with light quarks, they are expected to generate loop-induced contributions to baryon number violating processes. This was studied in the context of $\Delta B=2$ processes in \cite{ThomasArun:2025rgx}, while, in here, we extend the analysis to $\Delta B=1$ processes. In contrast to previous approach, here we evolve the Wilson coefficient of proton decay operator with the beta function appended by the contribution from BNC dimension-6 operators which mix with the BNV operators at 1-loop. A schematic of the analysis is given in Fig.~\ref{fig:block_diagram_rep}. The loop diagrams (b) and (c) correspond to mixing in proton decay and mixing in the neutron decay processes respectively. Note that the contribution from light first two generation quarks, in the loop, is numerically suppressed. Although at first sight this contribution may appear to be suppressed by $\frac{m_t^2}{\Lambda^4}$, the insufficient experimental data on flavour violating BNC operator of the form `$\bar{t}u\bar{t}u$' and BNV operators like `$ttde^+$' allow for potential significant effects. While the analysis in this work focuses on the $p\to e^+ \pi^0$ process, the framework can be effortlessly extended to other decay modes.

When all terms in the effective Lagrangian are taken into account, the BNV operators at dimension-8 are at the same order ($\mathcal{O}(\frac{1}{\Lambda^4})$) in expansion as the BNC-BNV mixing terms. Despite this similarity, their contributions to the RG evolution of the proton decay operator remains suppressed. To see this, recall that the dimension-8 BNV operators take the general forms~\cite{Murphy:2020rsh} $\psi^4 H^2$, $\psi^4 X$, $\psi^4 D^2$, and $\psi^4 H D$, where $H$, $X$, and $D$ denote the Higgs field, field-strength tensor, and covariant derivative, respectively. The contributions of these opeartors are either highly constrained, suppressed or appear at 2-loops compared to the BNC-BNV mixing effect. Nevertheless, they will induce other interesting processes such as $p \to e^+ e^+ e^-$, but this analysis lies beyond the scope of the present work.

To facilitate further studies, we provide a publicly available Python package~\cite{RepoName} that performs the RG evolution. This tool is designed to flexible and easily accommodate a wide class of BNC and BNV Standard Model Effective Field Theory (SMEFT) new-physics operators at different choices of threshold energy scales, enabling users to generalize our analysis to other nucleon decays and to any BSM scenario they might wish to explore. 

In Section.~\ref{sec:Theo_frame}, we briefly introduce the SMEFT framework and the dimension-6 baryon number conserving and baryon number violating operators. We also derive RG evolution of the proton decay operator, including the BNC-BNV mising at 1-loop. The running of the Wilson Coefficients are given in Section.~\ref{sec:BF-running}, where we consider three scenarios corresponding to three different scales of BNC new-physics namely, $10^4$ GeV, $10^6$ GeV and $10^9$ GeV. The proton deacy Wilson coefficient is then evolved from $10^3$ GeV to these BNC energy sales. In Section.~\ref{sec:python}, we introduce the Python code that we used in this analysis and finally in Section.~\ref{sec:Summary}, we summarise our results. For readability, all the notations used in the paper are consolidated in Appendix~\ref{app:index}.

\section{Theoretical Framework}\label{sec:Theo_frame}
In this section, we set up the theoretical framework required for our study. We begin with a brief overview of the SMEFT framework, with emphasis on dimension-6 BNC and BNV operators, highlighting the subset of operators that mix at 1-loop. These BNC-BNV mixing is later used to study the modification to RG evolution of Wilson coefficients of operators that generate proton decay.

\subsection{Standard Model Effective Field Theory}
\label{subsec:SMEFT}
In order to study the running of BNV within a systematic and model independent approach, we make use of the SMEFT framework. Here, the Standard Model (SM) Lagrangian is appended by higher-dimensional operators built out of SM fields, respecting the SM gauge symmetries, with each operator obtained by integrating out the non-dynamical new-physics at scale $\Lambda$~\cite{Nieves:1981tv,Buchmuller:1985jz,Grzadkowski:2010es,Jenkins:2013zja,Jenkins:2013wua,Falkowskiarticle,Lehman:2014jma,PhysRevLett.43.1571,Aoki:2016frl,RevModPhys.96.015006,Contino:2016jqw,PhysRevLett.43.1566,Helset:2019eyc,Murphy:2020rsh,Bonnefoy:2020tyv,PhysRevD.22.2208,Alonso:2013hga,DelDebbio:2013uaa,Alonso:2014rga, Wang:2023bdw,Jenkins:2017jig,Jenkins:2017dyc,Liao:2019tep,Zhang:2023ndw,Zhang:2023kvw,Biswas:2020abl}. The full Lagrangian can then be written as,
\begin{equation}
\mathcal{L}_{\text{SMEFT}} \;=\; \mathcal{L}_{\text{SM}} \;+\; \sum_{d>4} \frac{1}{\Lambda^{d-4}} \, \mathcal{L}^{(d)} ,
\end{equation}
where $\mathcal{L}_{\text{SM}}$ is the renormalisable Standard Model Lagrangian, and $\mathcal{L}^{(d)}$ contains operators of mass dimension $d$.

The Lagrangian at dimension-6 operators takes the form,
\begin{equation}
\mathcal{L}^{(6)} \;=\; \sum_i C_{i,prst} \, Q_{i,prst} ,
\end{equation}
with $Q_{i,prst}$ denoting the complete, independent set of dimension-6 operators and $c_{i,prst}$ their corresponding dimensionless Wilson coefficients. And $prst$, denotes the family indices of the four fermion fields. In our calculation we use dimentionful Wilson coefficients $C_{i,prst}$, 
\begin{equation}
\label{eq:smeftWC}
C_{i,prst} \;=\; \frac{c_{i,prst}}{\Lambda^2}.
\end{equation}

These dimension-6 operators have been classified into a complete and non-redundant basis. The lowest-dimensional BNV operators appear at dimension-6, where baryon number is violated by one unit ($\Delta B = 1$). These operators play a central role in nucleon decay [Table.~\ref{tab:decay_channels}] and related processes. 

\paragraph{Baryon Number Conserving Operators:}
The BNC sector of SMEFT at dimension-6 is remarkably rich. After using the equations of motion to eliminate redundant terms, there are 59 independent BNC operators. These operators are built exclusively from SM fields, preserve baryon and lepton number, and can be systematically grouped according to their field content:
\begin{itemize}
  \item Purely bosonic operators, involving only Higgs and gauge fields. Examples include $X^3$ operators (triple gauge field strengths), $H^6$ operators (six Higgs fields), $H^4 D^2$ operators (operators with four Higgs fields and two derivatives), and Higgs-gauge combinations such as $H^2 X^2$.
  \item Two-fermion operators, which modify fermion-Higgs interactions or generate dipole-like couplings. These include terms of the form $\psi^2 H^3$ (correcting Yukawa couplings), $\psi^2 H^2 D$ (modifying fermion currents), and $\psi^2 XH$ (dipole operators).
  \item Four-fermion operators, which introduce new contact interactions between fermions. These form the largest class and include combinations of left-left, right-right, and mixed chiralities.
\end{itemize}
Although the number of operator structures is only 59, when one accounts for flavour indices, the parameter space expands enormously. For three generations, the complete dimension-6 BNC Lagrangian contains 1350 \textit{CP}-even and 1149 \textit{CP}-odd parameters, yielding a total of 2499 real independent couplings. These parameters encode possible new-physics effects across a vast range of processes, including Higgs production and decay, electroweak precision observables, dipole transitions, and flavour physics.

\paragraph{Baryon Number Violating Operators:}
In the SMEFT framework, baryon number can be violated starting at dimension-6 with four-fermion interactions involving three quarks and one lepton appear. These operators change baryon number by one unit ($\Delta B = 1$), and lepton number by one unit ($\Delta L = 1$), such that $B-L$ remains conserved. They are the leading effective interactions responsible for nucleon decay in many ultraviolet completions of the SM, such as Grand Unified Theories (GUTs).

In the Warsaw basis, there are four independent $\Delta B = \Delta L = 1$ operators at dimension-6:
\begin{align}
Q_{prst}^{duq\ell} &= \epsilon_{\alpha\beta\gamma}\epsilon_{ij}(d^{\alpha}_{p}Cu^{\beta}_{r})(q^{i\gamma}_{s}C\ell^{j}_{t}), \\ \label{eq:op_duql}
Q_{prst}^{qque} &= \epsilon_{\alpha\beta\gamma}\epsilon_{ij}(q^{i\alpha}_{p}Cq^{j\beta}_{r})(u^{\gamma}_{s}Ce_{t}), \\
Q_{prst}^{qqq\ell} &= \epsilon_{\alpha\beta\gamma}\epsilon_{il}\epsilon_{jk}(q^{i\alpha}_{p}Cq^{j\beta}_{r})(q^{k\gamma}_{s}C\ell^{l}_{t}), \\
Q_{prst}^{duue} &= \epsilon_{\alpha\beta\gamma}(d^{\alpha}_{p}Cu^{\beta}_{r})(u^{\gamma}_{s}Ce_{t}), \label{eq:op_duue}
\end{align} 
where, $C$ denotes the charge conjugation operator. The fields $q$ and $\ell$ correspond to the left-handed quark and lepton doublets, while $u$, $d$, and $e$ represent the right-handed up-type quark, down-type quark, and charged lepton fields, respectively. Greek indices label $SU(3)_{c}$ color, Roman indices $i$-$l$ are used for $SU(2)_{L}$ components, and the letters $p,r,s,\text{and},t$ denote flavor (generation) indices taking values $1,\ldots,n_{g}=3$.

Altogether, the dimension-6 SMEFT Lagrangian can therefore be schematically written as,
\begin{equation}
\mathcal{L}^{(6)} \;=\; \sum_{i=1}^{59} C^{\text{BNC}}_{i,prst} \, Q^{\text{BNC}}_{i,prst} 
\;+\; \sum_{j=1}^{4} C^{\text{BNV}}_{j,prst} \, Q^{\text{BNV}}_{j,prst} \;=\; \sum_{k=1}^{63} C_{k,prst} \, Q_{k,prst},
\end{equation}
where the second sum ($j$) runs over the four BNV operators.

The Wilson coefficients $C_{k,prst}(\mu)$ run with the renormalisation scale $\mu$ according to
\begin{equation}
\dot C_{k,prst}(\mu) = 16\pi^2 \, \mu \frac{d C_{k,prst}}{d \mu} \;=\; \sum_l \gamma_{kl} \, C_{l,prst} ,
\end{equation}
where $\gamma_{kl}$ is the 1-loop anomalous dimension matrix.

The matching conditions for Wilson coefficients from the electro-weak scale (e.g.\ $10^3$~GeV) down to the hadronic scale are derived in the Ref.~\cite{Beneito:2023xbk}. Since in this work, our aim is to understand the involvement of dimension-6 BNC operators lying at an intermediate scale between the electro-weak scale and BNV new-physics scale, we will adopt matching below electro-weak scale as in previous literature and do not repeat the discussion.

\subsection{Loop Diagrams from BNC-BNV Mixing and Their Beta Functions}
\label{subsec::BNV-BNC}
\subsubsection{BNC-BNV Mixing}
\label{subsubsec:BNV-BNC-Mix}
At the 1-loop level, the situation becomes richer. Insertions of the BNV operators can mix with SMEFT operators of the type $(\bar{q}q)(\bar{q}q)$, i.e. dimension-6 four-fermion interactions, or with operators involving electroweak bosons and the Higgs sector. The ``BNV block'' in the Fig.~\ref{fig:block_diagram_rep} represents the proton decay operators, while the ``BNC block'' encapsulates the loop-induced contributions from the Standard Model fields and any possible intermediate new-physics degrees of freedom (for nucleon decay, these SMEFT contributions can only be BNC operators). Together, these blocks define the full EFT picture relevant for connecting high-scale baryon number violation with observable low-energy decay channels.

Among the complete set of 63 dimension-6 SMEFT operators, only 22 operators enter into the loop-level BNC-BNV mixing for nucleon decay. The restriction arises because the fermionic structure of the BNV operators requires that the insertion from the BNC sector provide the correct spinor, colour, and gauge contractions to form a consistent 1-loop diagram. Operators that are purely bosonic, or that involve mismatched chirality structures, cannot close the loop with a BNV vertex. 

The surviving set is therefore composed mainly of four-fermion operators of the $(\bar{L}L)(\bar{L}L)$, $(\bar{R}R)(\bar{R}R)$, $(\bar{L}L)(\bar{R}R)$, and mixed $(\bar{L}R)(\bar{R}L)$ types, together with a handful of $(\bar{L}R)(\bar{L}R)$ structures and their Hermitian conjugates. For reference, we list the full set of these 22 operators in Table.~\ref{tab:BNV-BNC-mixing}.

\begin{table}
\begin{center}
\small
\begin{minipage}[t]{4.75cm}
\renewcommand{\arraystretch}{1.5}
\begin{tabular}[t]{c|c}
\multicolumn{2}{c}{$(\bar LL)(\bar LL)$} \\
\hline
\hline
$Q_{qq}^{(1)}$  & $(\bar q_p \gamma_\mu q_r)(\bar q_s \gamma^\mu q_t)$ \\
$Q_{qq}^{(3)}$  & $(\bar q_p \gamma_\mu \tau^I q_r)(\bar q_s \gamma^\mu \tau^I q_t)$ \\
$Q_{lq}^{(1)}$  & $(\bar l_p \gamma_\mu l_r)(\bar q_s \gamma^\mu q_t)$ \\
$Q_{lq}^{(3)}$  & $(\bar l_p \gamma_\mu \tau^I l_r)(\bar q_s \gamma^\mu \tau^I q_t)$
\end{tabular}
\end{minipage}
\begin{minipage}[t]{5.25cm}
\renewcommand{\arraystretch}{1.5}
\begin{tabular}[t]{c|c}
\multicolumn{2}{c}{$(\bar RR)(\bar RR)$} \\
\hline
\hline
$Q_{uu}$ & $(\bar u_p \gamma_\mu u_r)(\bar u_s \gamma^\mu u_t)$ \\
$Q_{dd}$ & $(\bar d_p \gamma_\mu d_r)(\bar d_s \gamma^\mu d_t)$ \\
$Q_{eu}$ & $(\bar e_p \gamma_\mu e_r)(\bar u_s \gamma^\mu u_t)$ \\
$Q_{ed}$ & $(\bar e_p \gamma_\mu e_r)(\bar d_s\gamma^\mu d_t)$ \\
$Q_{ud}^{(1)}$ & $(\bar u_p \gamma_\mu u_r)(\bar d_s \gamma^\mu d_t)$ \\
$Q_{ud}^{(8)}$ & $(\bar u_p \gamma_\mu T^A u_r)(\bar d_s \gamma^\mu T^A d_t)$ \\
\end{tabular}
\end{minipage}
\begin{minipage}[t]{4.75cm}
\renewcommand{\arraystretch}{1.5}
\begin{tabular}[t]{c|c}
\multicolumn{2}{c}{$(\bar LL)(\bar RR)$} \\
\hline
\hline
$Q_{lu}$ & $(\bar l_p \gamma_\mu l_r)(\bar u_s \gamma^\mu u_t)$ \\
$Q_{ld}$ & $(\bar l_p \gamma_\mu l_r)(\bar d_s \gamma^\mu d_t)$ \\
$Q_{qe}$ & $(\bar q_p \gamma_\mu q_r)(\bar e_s \gamma^\mu e_t)$ \\
$Q_{qu}^{(1)}$ & $(\bar q_p \gamma_\mu q_r)(\bar u_s \gamma^\mu u_t)$ \\
$Q_{qu}^{(8)}$ & $(\bar q_p \gamma_\mu T^A q_r)(\bar u_s \gamma^\mu T^A u_t)$ \\
$Q_{qd}^{(1)}$ & $(\bar q_p \gamma_\mu q_r)(\bar d_s \gamma^\mu d_t)$ \\
$Q_{qd}^{(8)}$ & $(\bar q_p \gamma_\mu T^A q_r)(\bar d_s \gamma^\mu T^A d_t)$\\
\end{tabular}
\end{minipage}

\vspace{0.25cm}

\begin{minipage}[t]{3.75cm}
\renewcommand{\arraystretch}{1.5}
\begin{tabular}[t]{c|c}
\multicolumn{2}{c}{$(\bar LR)(\bar RL)+\hbox{h.c.}$} \\
\hline
\hline
$Q_{ledq}$ & $(\bar l_p^j e_r)(\bar d_s q_{tj})$
\end{tabular}
\end{minipage}
\begin{minipage}[t]{5.5cm}
\renewcommand{\arraystretch}{1.5}
\begin{tabular}[t]{c|c}
\multicolumn{2}{c}{$(\bar LR)(\bar L R)+\hbox{h.c.}$} \\
\hline
\hline
$Q_{quqd}^{(1)}$ & $(\bar q_p^j u_r) \epsilon_{jk} (\bar q_s^k d_t)$ \\
$Q_{quqd}^{(8)}$ & $(\bar q_p^j T^A u_r) \epsilon_{jk} (\bar q_s^k T^A d_t)$ \\
$Q_{lequ}^{(1)}$ & $(\bar l_p^j e_r) \epsilon_{jk} (\bar q_s^k u_t)$ \\
$Q_{lequ}^{(3)}$ & $(\bar l_p^j \sigma_{\mu\nu} e_r) \epsilon_{jk} (\bar q_s^k \sigma^{\mu\nu} u_t)$
\end{tabular}
\end{minipage}
\end{center}
\caption{\label{tab:BNV-BNC-mixing}
Baryon number conserving dimension-6 SMEFT operators (22 in total) that contribute to 1-loop mixing with baryon number violating operators. The subscripts $p,r,s,t$ are flavour indices.}
\end{table}

\subsubsection{Beta Function Calculation}
\label{subsubsec:Beta-calc}
With the identification of the 22 BNC operators that mix with BNV vertices, their 1-loop anomalous dimensions matrix could be found and the RG evolution for the BNV operators become,

\begin{eqnarray}\label{eq:duqRGE}
\dot C_{prst}^{duq\ell} &=& - C_{prst}^{duq\ell} \left[ 4 g_3^2 + \frac{9}{2} g_2^2 -6(\mathsf{y}_d \mathsf{y}_u + \mathsf{y}_q \mathsf{y}_l) g_1^2 \right] \nonumber \\
&&- C_{vrwt}^{duq\ell} (Y_d)_{vs} (Y_d^\dag)_{wp} - C_{pvwt}^{duq\ell} (Y_u)_{vs} (Y_u^\dag)_{wr} \nonumber \\
&& + \left\{ 2C_{prwv}^{duue} + C_{pwrv}^{duue} \right\} (Y_e)_{vt} (Y_u)_{ws} + 2C_{wsrv}^{qque} (Y_d^\dag)_{wp} (Y_e)_{vt} \nonumber \\
&& + \left\{ 2C_{vwst}^{qqq\ell} + 2C_{wvst}^{qqq\ell} - C_{vswt}^{qqq\ell} - C_{wsvt}^{qqq\ell} + 2C_{svwt}^{qqq\ell} + 2C_{swvt}^{qqq\ell} \right\} (Y_d^\dag)_{vp} (Y_u^\dag)_{wr}\nonumber \\
&& + C_{vrst}^{duq\ell} (Y_d Y_d^\dag)_{vp} + C_{pvst}^{duq\ell} (Y_u Y_u^\dag)_{vr} + \frac{1}{2} C_{prvt}^{duq\ell} (Y_u^\dag Y_u + Y_d^\dag Y_d)_{vs} \nonumber \\
&& + \frac{1}{2} C_{prsv}^{duq\ell} ( Y_e^\dag Y_e)_{vt}- \sum_{i}\sum_{j}a^{duq\ell}_{ij} C^{\text{BNC}}_{i,prst} \, C^{\text{BNV}}_{j,prst}
\end{eqnarray}
\begin{eqnarray}\label{eq:qqueRGE}
\dot C_{prst}^{qque} &=&- C_{prst}^{qque} \left[ 4 g_3^2 + \frac{9}{2} g_2^2 -6(\mathsf{y}_q^2 + \mathsf{y}_u \mathsf{y}_e) g_1^2 \right] \nonumber \\
&&- C_{pwvt}^{qque} (Y_u)_{vr} (Y_u^\dag)_{ws} - C_{rwvt}^{qque} (Y_u)_{vp} (Y_u^\dag)_{ws} \nonumber \\
&& + \frac{1}{2} C_{vspw}^{duq\ell} (Y_e^\dag)_{wt} (Y_d)_{vr} + \frac{1}{2} C_{vsrw}^{duq\ell} (Y_e^\dag)_{wt} (Y_d)_{vp} \nonumber \\
&& - \frac{1}{2} \left\{ 2 C_{vwst}^{duue} + C_{vswt}^{duue} \right\} \left[ (Y_d)_{vp} (Y_u)_{wr} + (Y_d)_{vr} (Y_u)_{wp} \right] \nonumber \\
&& + \frac{1}{2} \left\{ -2C_{prwv}^{qqq\ell} - 2C_{rpwv}^{qqq\ell} + C_{pwrv}^{qqq\ell} + C_{rwpv}^{qqq\ell} -2C_{wprv}^{qqq\ell} -2C_{wrpv}^{qqq\ell} \right\} (Y_u^\dag)_{ws} (Y_e^\dag)_{vt}\nonumber \\
&& + \frac{1}{2} C_{vrst}^{qque} (Y_u^\dag Y_u + Y_d^\dag Y_d)_{vp} + \frac{1}{2} C_{pvst}^{qque} (Y_u^\dag Y_u + Y_d^\dag Y_d)_{vr} + C_{prvt}^{qque} (Y_u Y_u^\dag)_{vs}\nonumber \\
&& + C_{prsv}^{qque} (Y_e Y_e^\dag)_{vt} - \sum_{i}\sum_{j}a^{qque}_{ij} C^{\text{BNC}}_{i,prst} \, C^{\text{BNV}}_{j,prst} 
\end{eqnarray}
\begin{eqnarray}\label{eq:qqqRGE}
\dot C_{prst}^{qqq\ell} &=& - C_{prst}^{qqq\ell} \left[ 4 g_3^2 + 3 g_2^2 -6(\mathsf{y}_q ^2 + \mathsf{y}_q \mathsf{y}_l) g_1^2 \right] \nonumber \\
&& - 4 \left\{ C_{rpst}^{qqq\ell} + C_{srpt}^{qqq\ell} + C_{psrt}^{qqq\ell} \right\} g_2^2 - 4C_{prwv}^{qque} (Y_e)_{vt} (Y_u)_{ws}\nonumber \\
&&  + 2C_{vwst}^{duq\ell} \left[ (Y_d)_{vp} (Y_u)_{wr} + (Y_d)_{vr} (Y_u)_{wp} \right]\nonumber \\
&&  + \frac{1}{2} C_{vrst}^{qqq\ell} (Y_u^\dag Y_u + Y_d^\dag Y_d)_{vp} + \frac{1}{2} C_{pvst}^{qqq\ell} (Y_u^\dag Y_u + Y_d^\dag Y_d)_{vr}\nonumber \\
&& + \frac{1}{2} C_{prvt}^{qqq\ell} (Y_u^\dag Y_u + Y_d^\dag Y_d)_{vs} + \frac{1}{2} C_{prsv}^{qqq\ell} ( Y_e^\dag Y_e)_{vt} - \sum_{i}\sum_{j}a^{qqql}_{ij} C^{\text{BNC}}_{i,prst} \, C^{\text{BNV}}_{j,prst}
\end{eqnarray}
\begin{eqnarray}\label{eq:duuRGE}
\dot C_{prst}^{duue} &=& - C_{prst}^{duue} \left[ 4 g_3^2 -2 \left( 2 \mathsf{y}_d \mathsf{y}_u +2 \mathsf{y}_e \mathsf{y}_u + \mathsf{y}_u^2+ \mathsf{y}_e\mathsf{y}_d \right) g_1^2 \right] \nonumber \\
&&+ 4C_{psrt}^{duue}\left((\mathsf{y}_d +\mathsf{y}_e)\mathsf{y}_u - \mathsf{y}_u^2- \mathsf{y}_e\mathsf{y}_d \right)g_1^2 \nonumber \\
&&+ 4C_{prwv}^{duq\ell} (Y_u^\dag)_{ws} (Y_e^\dag)_{vt} - 8C_{vwst}^{qque} (Y_d^\dag)_{vp} (Y_u^\dag)_{wr} + C_{vrst}^{duue} (Y_d Y_d^\dag)_{vp}\nonumber \\
&& + C_{pvst}^{duue} (Y_u Y_u^\dag)_{vr} + C_{prvt}^{duue} (Y_u Y_u^\dag)_{vs} + C_{prsv}^{duue} (Y_e Y_e^\dag)_{vt} - \sum_{i}\sum_{j}a^{duue}_{ij} C^{\text{BNC}}_{i,prst} \, C^{\text{BNV}}_{j,prst}
\end{eqnarray}

Where $g_1$, $g_2$, and $g_3$ are the gauge couplings of the Standard Model corresponding to the $U(1)_Y$ hypercharge, $SU(2)_L$ weak isospin, and $SU(3)_c$ colour interactions, respectively. The matrices $Y_u$, $Y_d$, and $Y_e$ denote the Yukawa couplings for up-type quarks, down-type quarks, and charged leptons, which enter through fermion-Higgs interactions and induce flavour-dependent operator mixing, and the subscripts $p,r,s,t,v,w$ are flavour indices. The coefficients $C^{\text{BNV}}_{i,prst}$ represent the Wilson coefficients of the dimension-6 BNV operators, while $C^{\text{BNC}}_{j,prst}$ are those of the BNC SMEFT operators. Finally, $a_{ij}$ are the mixing coefficients introduced to parameterise the 1-loop contributions that connect the BNC and BNV sectors, extending the usual RG equations in which BNV operators only mix among themselves.

\begin{figure}[h!]
    \centering
    \includegraphics[scale=0.17]{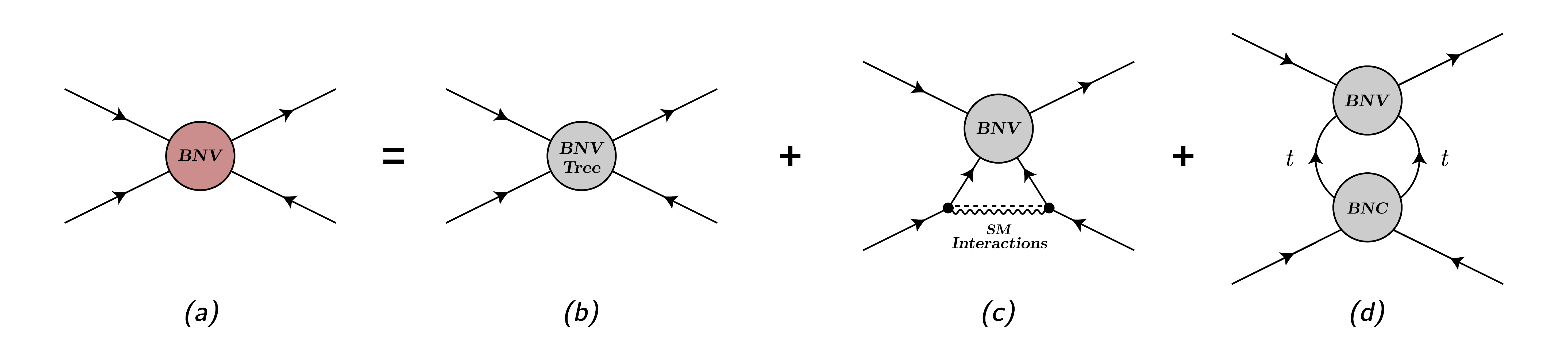}
    \caption{Diagrammatic representation of BNV running (a) consists of tree (b), the SM interaction (c), and the BNC-BNV  mixing (d) of first-generation BNV operators. The dominant loop contribution BNC-BNV  mixing arises from top-top contractions, which feed into the anomalous dimensions of the BNV coefficients.}
    \label{fig:BNV-BNC-loops}
\end{figure}

At tree level, the first-generation BNV operators directly mediate the decay $p \to \pi^0 e^+$, which provides the leading experimental constraint on their Wilson coefficients~\cite{Beneito:2023xbk,Super-Kamiokande:2020wjk}. When loop effects are included, these same operators mix with BNC insertions through diagrams in which heavy quarks propagate in the loop. Among the possible choices, the top-top contractions give the dominant contribution to the anomalous dimensions, as illustrated in Fig.~\ref{fig:BNV-BNC-loops}, which shows the Standard Model interaction together with the BNC-BNV mixing of the first-generation BNV operators. This implies that the most stringent bounds on the first-generation BNV coefficients receive loop-induced corrections controlled by the BNC sector, with the mixing coefficients $a_{ij}$ quantifying the strength of this connection. The explicit mapping between each proton decay BNV operator, the BNC and BNV operators with flavour violating structure that can mix into it, and the corresponding 1-loop coefficient is summarised in Table.~\ref{tab:mixing_terms}. For completeness, the full calculation of a representative 1-loop diagram is presented in Appendix~\ref{app:loop-calc}.
\begin{table}[h]
\centering
\begin{tabular}{c|c|c|c}
\hline
\hline
Proton decay & BNV operator contributing & BNC operator contributing & \multirow{2}{*}{$a_{ij}$} \\
operator & in the 1-loop ($C^{\text{BNV}}_{i,prst}$) & in the 1-loop ($C^{\text{BNC}}_{j,prst}$) &  \\
\hline
\hline \rule{0pt}{4ex}
\multirow{2}{*}{$Q^{duql}_{1111}$}  & $C^{duql}_{1331}$ & $C^{qu1}_{1313}$ & $\dfrac{-12}{16\pi^2}m^2_t$  \\[8pt]
\cdashline{2-4}\rule{0pt}{4ex}
& $C^{duql}_{1331}$ & $C^{qu8}_{1313}$ & $\dfrac{-48}{16\pi^2}m^2_t$  \\[8pt]
\hline\rule{0pt}{4ex}
\multirow{2}{*}{$Q^{qqql}_{1111}$}  & $C^{qqql}_{1331}$ & $C^{qq1}_{1313}$ & $\dfrac{-96}{16\pi^2}m^2_t$  \\[8pt]
\cdashline{2-4}\rule{0pt}{4ex}
& $C^{qqql}_{1331}$ & $C^{qq3}_{1313}$ & $\dfrac{-576}{16\pi^2}m^2_t$  \\[8pt]
\hline\rule{0pt}{4ex}
\multirow{2}{*}{$Q^{qque}_{1111}$}  & $C^{qque}_{1331}$ & $C^{qu1}_{1313}$ & $\dfrac{-24}{16\pi^2}m^2_t$ \\[8pt]
\cdashline{2-4}\rule{0pt}{4ex}
& $C^{qque}_{1331}$ & $C^{qu8}_{1313}$ & $\dfrac{-96}{16\pi^2}m^2_t$ \\[8pt]
\hline\rule{0pt}{4ex}
$Q^{duue}_{1111}$  & $C^{duue}_{1331}$ & $C^{uu}_{1313}$ & $\dfrac{-48}{16\pi^2}m^2_t$  \\[8pt]
\hline
\hline
\end{tabular}
\caption{Loop-induced mixing coefficients $a_{ij}$ entering the last term of Eqs.~(\ref{eq:duqRGE})-(\ref{eq:duuRGE}), $\sum_{i}\sum_{j} a_{ij}\,C^{\text{BNC}}_{i,prst}\,C^{\text{BNV}}_{j,prst}$, relevant for the decay $p \to \pi^0 e^+$. The dominant contributions arise from top-top contractions in the loop, with $m_t$ denoting the top quark mass (Appendix~\ref{app:index}).}
\label{tab:mixing_terms}
\end{table}

\section{Running of the Beta Function}\label{sec:BF-running}
In this section, we explore the scale dependence of the beta function through numerical running. The numerical framework allows us to evolve the Wilson coefficients and coupling parameters with energy, thereby illustrating how the coupled RG equations govern the behaviour of the effective couplings associated with BNV and BNC operators.

The running is performed for the scale \(\Lambda / \sqrt{c}\), where \(\Lambda\) represents the new-physics scale suppression and `\(c\)' denotes the dimensionless Wilson coefficient of the effective operator defined in Eq.~(\ref{eq:smeftWC}). This quantity serves as a convenient parameterisation of the effective interaction strength at different energy scales. Using the computed 1-loop beta functions from the previous section, the Wilson coefficients are evolved from $10^3$ GeV, electro-weak scale upward to the SMEFT scale, incorporating the full 1-loop contributions from both BNC and BNV sectors. Since the baryon number conserving (BNC) new-physics operator of the form $\bar{t}u\bar{t}u$ is not constrained to scales above $10^4~\text{GeV}$ by current Large Hadron Collider (LHC) data, we consider three representative scenarios with the BNC scale set at $10^4~\text{GeV}$, $10^6~\text{GeV}$, and $10^9~\text{GeV}$ for a comparative analysis. The initial values for both the BNC and BNV Wilson coefficients used in the analysis are summarised in Table.~\ref{tab:initial_conditions}.


\begin{table}[h!]
\centering
\renewcommand{\arraystretch}{1.5} 
\setlength{\tabcolsep}{10pt}
\begin{tabular}{c|c|c|c|c}
\hline
\hline
\multirow{2}{*}{Operator} & \multirow{2}{*}{ Set of Initial} & \multicolumn{3}{c}{Wilson coefficient [in GeV$^{-2}$] values at $\mu = 10^3$ GeV}  \\ \cline{3-5}
& Scales & for Case 1 & for Case 2 & for Case 3 \\
\hline
\hline
$Q^{duql}_{1111}$ &  & \multicolumn{3}{c}{$6.57462\times10^{-32}$~\cite{Beneito:2023xbk}} \\ 
\hline
$Q^{qque}_{1111}$ &  & \multicolumn{3}{c}{$3.30579\times10^{-32}$~\cite{Beneito:2023xbk}}\\ 
\hline
$Q^{qqql}_{1111}$ &  & \multicolumn{3}{c}{$6.57462\times10^{-32}$~\cite{Beneito:2023xbk}} \\ 
\hline
$Q^{duue}_{1111}$ &  & \multicolumn{3}{c}{$6.57462\times10^{-32}$~\cite{Beneito:2023xbk}} \\ 
\hdashline
$Q^{qu1}_{1313}$ & & $2\times10^{-8}$~\cite{ATLAS:2024hac} & $10^{-12}$ & $10^{-18}$ \\ 
\hline
$Q^{qu8}_{1313}$ & & $4.1\times10^{-8}$~\cite{ATLAS:2024hac} & $10^{-12}$ & $10^{-18}$ \\ 
\hline
$Q^{qq1}_{1313}$ & & $1.2\times10^{-7}$~\cite{Malito:2887295} & $10^{-12}$ & $10^{-18}$ \\ 
\hline
$Q^{qq3}_{1313}$ & & $1.2\times10^{-7}$~\cite{Malito:2887295} & $10^{-12}$ & $10^{-18}$ \\  
\hline
$Q^{uu}_{1313}$ & & $6.8\times10^{-9}$~\cite{ATLAS:2024hac} & $10^{-12}$ & $10^{-18}$ \\ 
\hdashline
\multirow{2}{*}{$Q^{i}_{1331}$} & Set 1 & $10^{-10}$ & $10^{-14}$ & $4\times10^{-20}$ \\ \cline{2-5}
\multirow{2}{*}{$i \in \{duql,\, qque, $} & Set 2 & $10^{-16}$ & $10^{-18}$ & $10^{-20}$ \\ \cline{2-5}
\multirow{2}{*}{$ \qquad qqql,\, duue \}$} & Set 3 & $10^{-22}$ & $10^{-22}$ & $4\times10^{-22}$ \\ \cline{2-5}
& Set 4 & $10^{-28}$ & $10^{-26}$ & $10^{-22}$ \\
\hline
\hline
\end{tabular}
\caption{Initial scale values for the Wilson coefficients (WCs) of baryon number violating (BNV) and baryon number conserving (BNC) operators at the reference scale of $10^3$~GeV, and their corresponding running scenarios defined as Case~1 (Sec.~\ref{subsec:BF-run-104}), Case~2 (Sec.~\ref{subsec:BF-run-106}), and Case~3 (Sec.~\ref{subsec:BF-run-109}). The operators above the first dashed line correspond to the first-generation BNV operators. The operators between the first and second dashed lines represent the BNC operators induced via operator mixing, while those below the second dashed line correspond to the third-generation BNV operators ($Q^{i}_{1331}$, with $i \in \{duql,\, qque,\, qqql,\, duue\}$) evaluated for four distinct sets of initial scale values.}

\label{tab:initial_conditions}
\end{table}

\subsection{Case 1: BNC operator scale at $10^4$ GeV}\label{subsec:BF-run-104}
The benchmark case corresponds to the experimentally observed upper limit on baryon number conserving (BNC) operators derived from top quark pair production in proton-proton collisions at \(\sqrt{s} = 10^3~\text{GeV}\) with the ATLAS detector~\cite{ATLAS:2024hac,Malito:2887295}. The initial scale values for the Wilson coefficients at the reference scale of $10^3$~GeV for the four baryon number violating operators $Q^{duql}_{1111}$, $Q^{qque}_{1111}$, $Q^{qqql}_{1111}$, and $Q^{duue}_{1111}$ are adopted from Ref.~\cite{ATLAS:2024hac}. Since the third-generation Wilson coefficients $C^{duql}_{1331}$, $C^{qque}_{1331}$, $C^{qqql}_{1331}$, and $C^{duue}_{1331}$ have not been observed by current experiments, we perform the running including the full BNC-BNV operator mixing, starting from distinct initial scale values for the third-generation coefficients. The corresponding initial values for both the BNV operators and their BNC mixing counterparts used in this case are summarised in Table.~\ref{tab:initial_conditions}.

\begin{figure}[htbp!]
    \centering
\hspace*{-0.5cm}\includegraphics[scale=0.55]{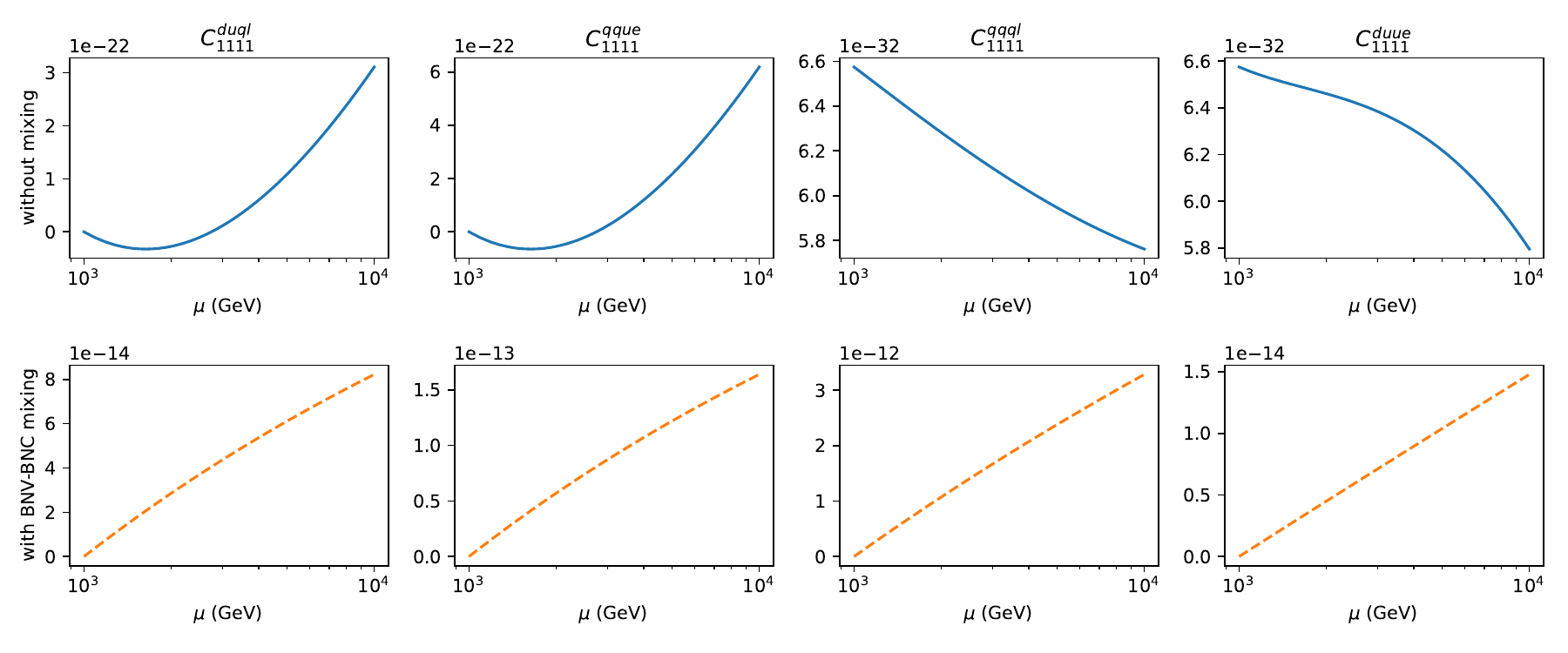}
    \caption{Running of the proton decay Wilson coefficients $C^{duql}_{1111}$, $C^{qque}_{1111}$, $C^{qqql}_{1111}$, and $C^{duue}_{1111}$ from $10^3$~GeV to $10^{4}~\text{GeV}$ as a function of the energy scale $\mu$. Solid blue lines show evolution without BNC interactions, while dotted orange lines include full BNC-BNV mixing. This running corresponds to the case with an initial value of $10^{-10}~\text{GeV}^{-2}$ (Set~1) for the Wilson coefficients $C^{duql}_{1331}$, $C^{qque}_{1331}$, $C^{qqql}_{1331}$, and $C^{duue}_{1331}$.}
    \label{fig:Case1_running}
\end{figure}

The running of the proton decay Wilson coefficients, $C^{duql}_{1111}$, $C^{qque}_{1111}$, $C^{qqql}_{1111}$, and $C^{duue}_{1111}$, as a function of the energy scale $\mu$ from $10^3$~GeV to $10^4$ GeV, for the initial value given by Set 1 in Table.~\ref{tab:initial_conditions}, are shown in Fig.~\ref{fig:Case1_running}. The results are presented for running with only the SM interactions (blue lines), and running with the BNC-BNV operator mixing contributions (dotted orange lines). It is observed that the sets of initial conditions used for the Wilson coefficients $C^{duql}_{1331}$, $C^{qque}_{1331}$, $C^{qqql}_{1331}$, and $C^{duue}_{1331}$ modifies the running with different significances.

\begin{figure}[ht!]
    \centering
\hspace*{-1.5cm}\includegraphics[scale=0.7]{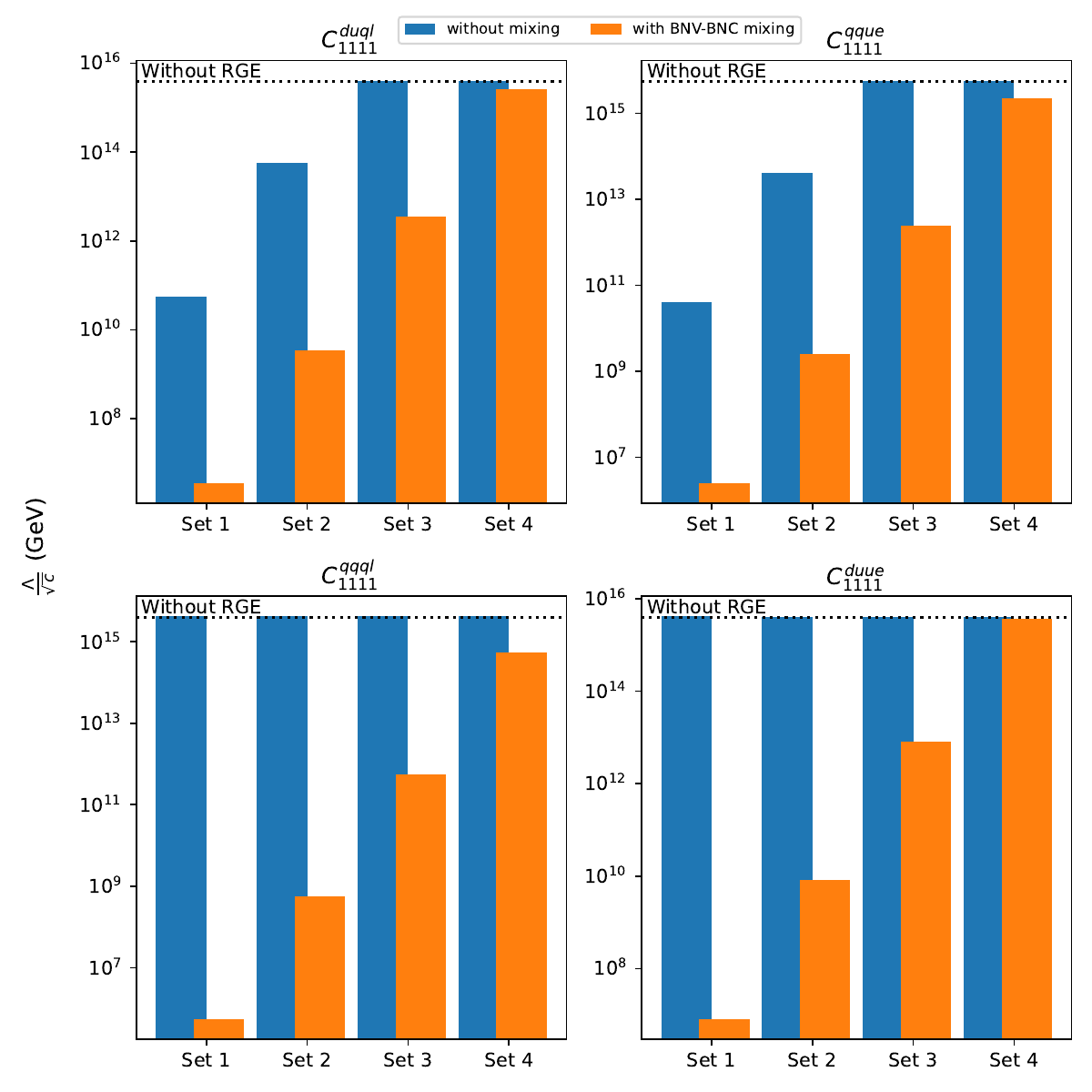}
    \caption{Comparison of the UV scale $\Lambda / \sqrt{c}$ for the four proton-decay operators in Eqs.~(\ref{eq:op_duql}-\ref{eq:op_duue}), contributing to the decay channel $p \to \pi^0 e^+$ for case 1. The dotted line represents without RG running and the two different colour bars correspond to: (i) RG running without BNC-BNV mixing interactions (in blue) and (ii) RG running including BNC-BNV operator mixing (in orange), for different initial scale values (Sets) of each BNV Wilson coefficient. Results are computed using the beta functions in Eqs.~(\ref{eq:duqRGE}-\ref{eq:duuRGE}), contributing to the decay channel $p \to \pi^0 e^+$ with the running scale $\mu$ from $10^3$~GeV to $10^4$ GeV.}
    \label{fig:Lambda_comparison_case1}
\end{figure}

\begin{figure}[htbp!]
\centering
    \begin{subfigure}{1\textwidth}
\centering
        \includegraphics[width=16cm,height=10.5cm]{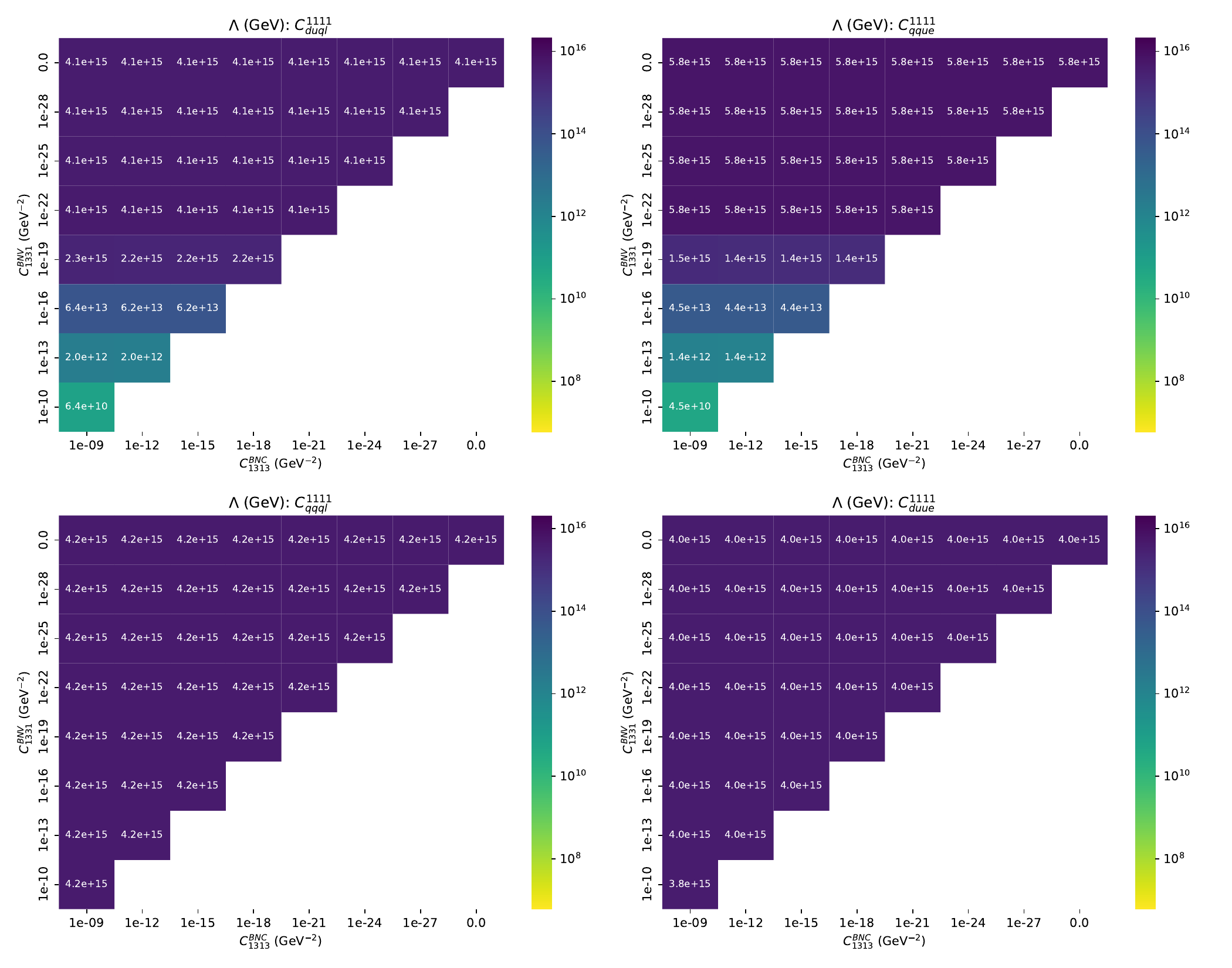}
        \caption{without BNC-BNV mixing terms}
\label{fig:without_1}
    \end{subfigure}
    \begin{subfigure}{1\textwidth}
\centering
        \includegraphics[width=16cm,height=10.5cm]{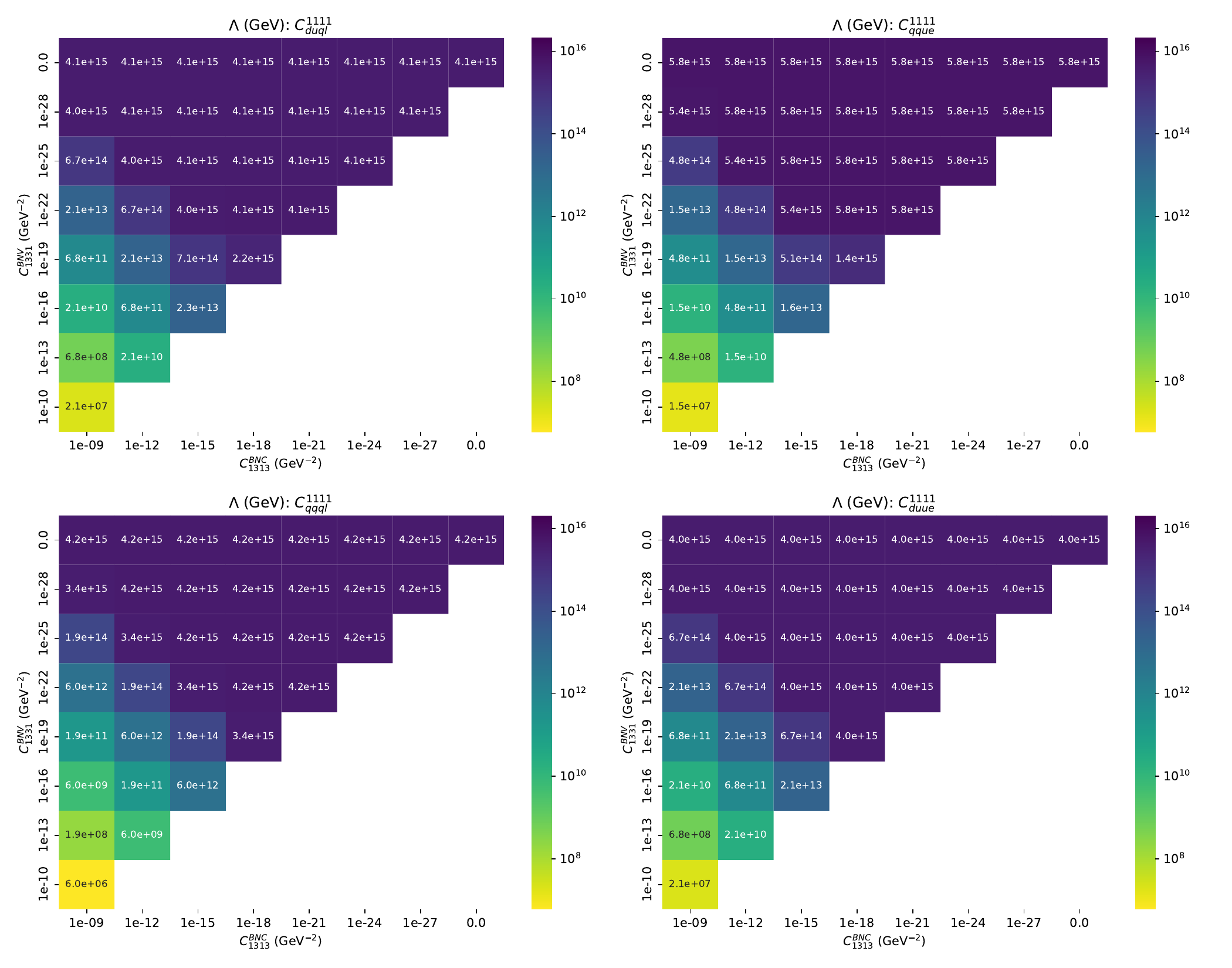}
        \caption{with BNC-BNV mixing terms}
\label{fig:with_1}
    \end{subfigure}
    \caption{The plot shows the dependence of the proton decay operator scale on the BNC-BNV parameter space for case 1. The x-axis represents the BNC operator Wilson coefficient ($C^{BNC}_{1313}$), while, the y-axis represent the BNV Wilson coefficient ($C^{BNV}_{1331}$).}
\label{fig:correlation_1}
\end{figure}

In Fig.~\ref{fig:Lambda_comparison_case1}, we present the comparison between the computed proton decay scale, $\Lambda / \sqrt{c}$, obtained for cases with no RG running (in dotted line), with RG running without including BNC interactions (in blue), and with RG running including both SM and BNC-BNV mixing (in orange), for all initial sets given in Table.~\ref{tab:initial_conditions}. These results use the RG evolutions defined in Eqs.~(\ref{eq:duqRGE}-\ref{eq:duuRGE}), that contribute to the proton decay channel $p \to \pi^0 e^+$. The comparison clearly demonstrates that the inclusion of RG effects, particularly the BNC-BNV mixing terms, have significant effect. The numerical values are given in Table.~\ref{tab:limits_case_1}. The dependence of the proton decay scale on the BNC-BNV mixing terms is summarised in Fig.~\ref{fig:correlation_1}. The plot Fig.\ref{fig:without_1} shows the proton decay scale obtained after RG evolution without including the BNC-BNV mixing, while Fig.~\ref{fig:with_1} shows the effect of the mixing.

\begin{table}[ht!]
\centering
\resizebox{\textwidth}{!}{%
\begin{tabular}{l|c|c|c|c|c|c|c|c}
\hline
\hline
BNV Operator & \multicolumn{4}{c|}{$Q^{duql}_{1111}$} & \multicolumn{4}{c}{$Q^{qque}_{1111}$}  \\
\hline
\hline
Set of Initial Scale & Set 1 & Set 2 & Set 3 & Set 4  & Set 1 & Set 2 & Set 3 & Set 4   \\
\hline
$\Lambda^{\mathrm{w/o}}/\sqrt{c}$ [in GeV] & \multicolumn{4}{c|}{$3.9\times10^{15}$} & \multicolumn{4}{c}{$5.5\times10^{15}$}  \\
\hline
\multirow{2}{*}{$\Lambda^{\mathrm{w/SM}}/\sqrt{c}$ [in GeV]} & $5.67$ & $5.67$ & $4.06$ & $4.07$ & $4.02$ & $4.02$ & $5.70$ & $5.76$  \\
& $\times10^{10}$ & $\times10^{13}$ & $\times10^{15}$ & $\times10^{15}$ & $\times10^{10}$ & $\times10^{13}$ & $\times10^{15}$ & $\times10^{15}$  \\
\hline
\multirow{2}{*}{$\Lambda^{\mathrm{w/mix}}/\sqrt{c}$ [in GeV]} & $3.49$ & $3.49$ & $3.49$ & $2.65$ & $2.47$ & $2.47$ & $2.47$ & $2.27$  \\
& $\times10^{6}$ & $\times10^{9}$ & $\times10^{12}$ & $\times10^{15}$ & $\times10^{6}$ & $\times10^{9}$ & $\times10^{12}$ & $\times10^{15}$  \\
\hline
\multirow{2}{*}{$\Lambda^{\mathrm{w/SM}}/\Lambda^{\mathrm{w/o}}$ } & $1.5$ & $1.5$ & \multirow{2}{*}{$1$} & \multirow{2}{*}{$1$} & $7.3$ & $7.3$ & \multirow{2}{*}{$1$} & \multirow{2}{*}{$1$} \\
& $\times10^{-5}$ & $\times10^{-2}$ &  &  & $\times10^{-6}$ & $\times10^{-3}$ & &   \\
\hline
\multirow{2}{*}{$\Lambda^{\mathrm{w/mix}}/\Lambda^{\mathrm{w/o}}$ }  & $8.9$ & $8.9$ & $8.9$ & \multirow{2}{*}{$0.68$} & $4.5$ & $4.5$ & $4.5$ & \multirow{2}{*}{$0.41$}   \\
& $\times10^{-10}$ & $\times10^{-7}$ & $\times10^{-4}$ &  &   $\times10^{-10}$ & $\times10^{-7}$ & $\times10^{-4}$ &   \\
\hline
\hline
BNV Operator &  \multicolumn{4}{c|}{$Q^{qqql}_{1111}$} & \multicolumn{4}{c}{$Q^{duue}_{1111}$}   \\
\hline
\hline
Set of Initial Scale &  Set 1 & Set 2 & Set 3 & Set 4  & Set 1 & Set 2 & Set 3 & Set 4  \\
\hline
$\Lambda^{\mathrm{w/o}}/\sqrt{c}$ [in GeV]  & \multicolumn{4}{c|}{$3.9\times10^{15}$} & \multicolumn{4}{c}{$3.9\times10^{15}$} \\
\hline
\multirow{2}{*}{$\Lambda^{\mathrm{w/SM}}/\sqrt{c}$ [in GeV]} & $4.17$ & $4.19$ & $4.19$ & $4.19$ & $4.15$ & $4.05$ & $4.05$ & $4.05$  \\
& $\times10^{15}$ & $\times10^{15}$ & $\times10^{15}$ & $\times10^{15}$ & $\times10^{15}$ & $\times10^{15}$ & $\times10^{15}$ & $\times10^{15}$  \\
\hline
\multirow{2}{*}{$\Lambda^{\mathrm{w/mix}}/\sqrt{c}$ [in GeV]}  & $5.52$ & $5.52$ & $5.52$ & $5.47$ & $8.22$ & $8.22$ & $8.22$ & $3.63$  \\
& $\times10^{5}$ & $\times10^{8}$ & $\times10^{11}$ & $\times10^{14}$ & $\times10^{6}$ & $\times10^{9}$ & $\times10^{12}$ & $\times10^{15}$ \\
\hline
$\Lambda^{\mathrm{w/SM}}/\Lambda^{\mathrm{w/o}}$ & $1.1$ & $1.1$ & $1.1$ & $1.1$ & $1.1$ & $1.0$ & $1.0$ & $1.0$  \\
\hline
\multirow{2}{*}{$\Lambda^{\mathrm{w/mix}}/\Lambda^{\mathrm{w/o}}$ }  & $1.4$ & $1.4$ & $1.4$ & \multirow{2}{*}{$0.14$} & $2.1$ & $2.1$ & $2.1$ & \multirow{2}{*}{$0.93$} \\
&  $\times10^{-10}$ & $\times10^{-7}$ & $\times10^{-4}$ &  &   $\times10^{-9}$ & $\times10^{-6}$ & $\times10^{-3}$ &     \\
\hline
\hline
\end{tabular}
}
\caption{Values of the UV scale $\Lambda/\sqrt{c}$ obtained under 4 sets of initial,  for case 1, considering without RG evolution ($\Lambda^\mathrm{w/o}/\sqrt{c}$), with RG running without including BNC effects ($\Lambda^\mathrm{w/SM}/\sqrt{c}$), and with RG running including both SM and BNC-BNV mixing ($\Lambda^\mathrm{w/mix}/\sqrt{c}$). The last two rows show the ratios $\Lambda^\mathrm{w/SM}/\Lambda^\mathrm{w/o}$ and $\Lambda^\mathrm{w/mix}/\Lambda^\mathrm{w/o}$.}
\label{tab:limits_case_1}
\end{table}

\subsection{Case 2: BNC operator scale at $10^6$ GeV}\label{subsec:BF-run-106}
In the second benchmark case, the limit on baryon number conserving (BNC) operators is assumed to be lower than the current experimental upper bound derived from top quark pair production at $\sqrt{s} = 10^3~\text{GeV}$. The running of the Wilson coefficients is performed from $10^3$~GeV up to $10^{6}~\text{GeV}$, with the same operator basis and initial scale values for the first- and third-generation coefficients as in Sec.~\ref{subsec:BF-run-104}. The initial values corresponding to this scenario are summarised in Table.~\ref{tab:initial_conditions}.

\begin{figure}[htbp!]
    \centering
\hspace*{-0.50cm}\includegraphics[scale=0.55]{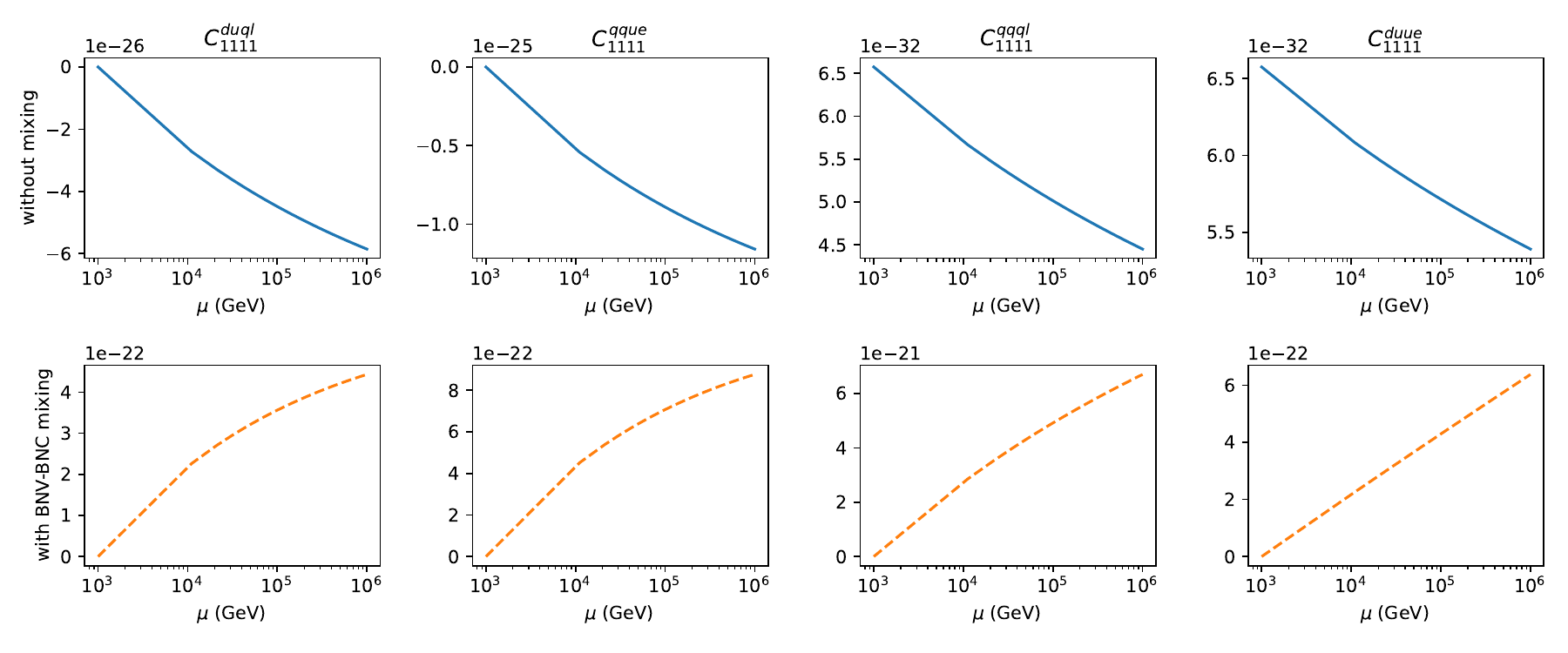}
    \caption{Evolution of the proton decay Wilson coefficients as in Fig.~\ref{fig:Case1_running}, where the running is performed as a function of the energy scale $\mu$ between $10^3$~GeV and $10^{6}$~GeV. This running corresponds to the case with an initial value of $10^{-14}~\text{GeV}^{-2}$ (Set~1) for the Wilson coefficients $C^{duql}_{1331}$, $C^{qque}_{1331}$, $C^{qqql}_{1331}$, and $C^{duue}_{1331}$.}
    \label{fig:Case2_running}
\end{figure}

Fig.~\ref{fig:Case2_running} shows the evolution of the four BNV Wilson coefficients $C^{duql}_{1111}$, $C^{qque}_{1111}$, $C^{qqql}_{1111}$, and $C^{duue}_{1111}$ over the energy range for scenario without mixing (blue) and full BNC-BNV mixing (dotted orange). Among the initial scale values, the benchmark with $C^{duql}_{1331} = C^{qque}_{1331} = C^{qqql}_{1331} = C^{duue}_{1331} = 10^{-14}~\text{GeV}^{-2}$ (Set 1) shows the most noticeable difference relative to the evolution.

\begin{figure}[ht!]
    \centering
\hspace*{-1.5cm}\includegraphics[scale=0.7]{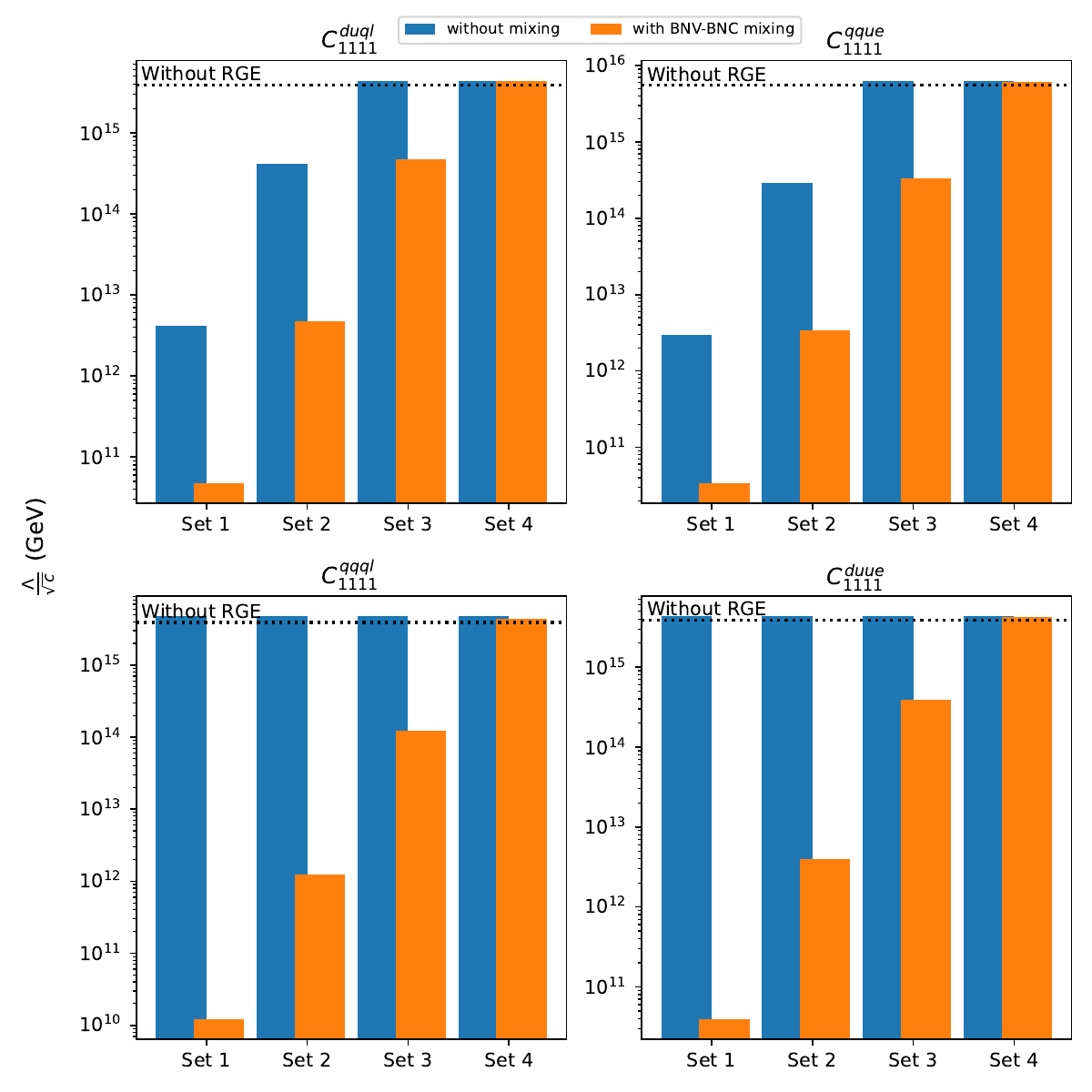}
    \caption{Comparison of the UV scale $\Lambda / \sqrt{c}$ as in Fig.~\ref{fig:Lambda_comparison_case1}, where the the coefficients are RG evolved with energy scale $\mu$ from $10^3$~GeV to $10^{6}$~GeV, for case 2.}
    \label{fig:Lambda_comparison_case2}
\end{figure}

\begin{figure}[htbp!]
\centering
    \begin{subfigure}{1\textwidth}
\centering
        \includegraphics[width=16cm,height=10.5cm]{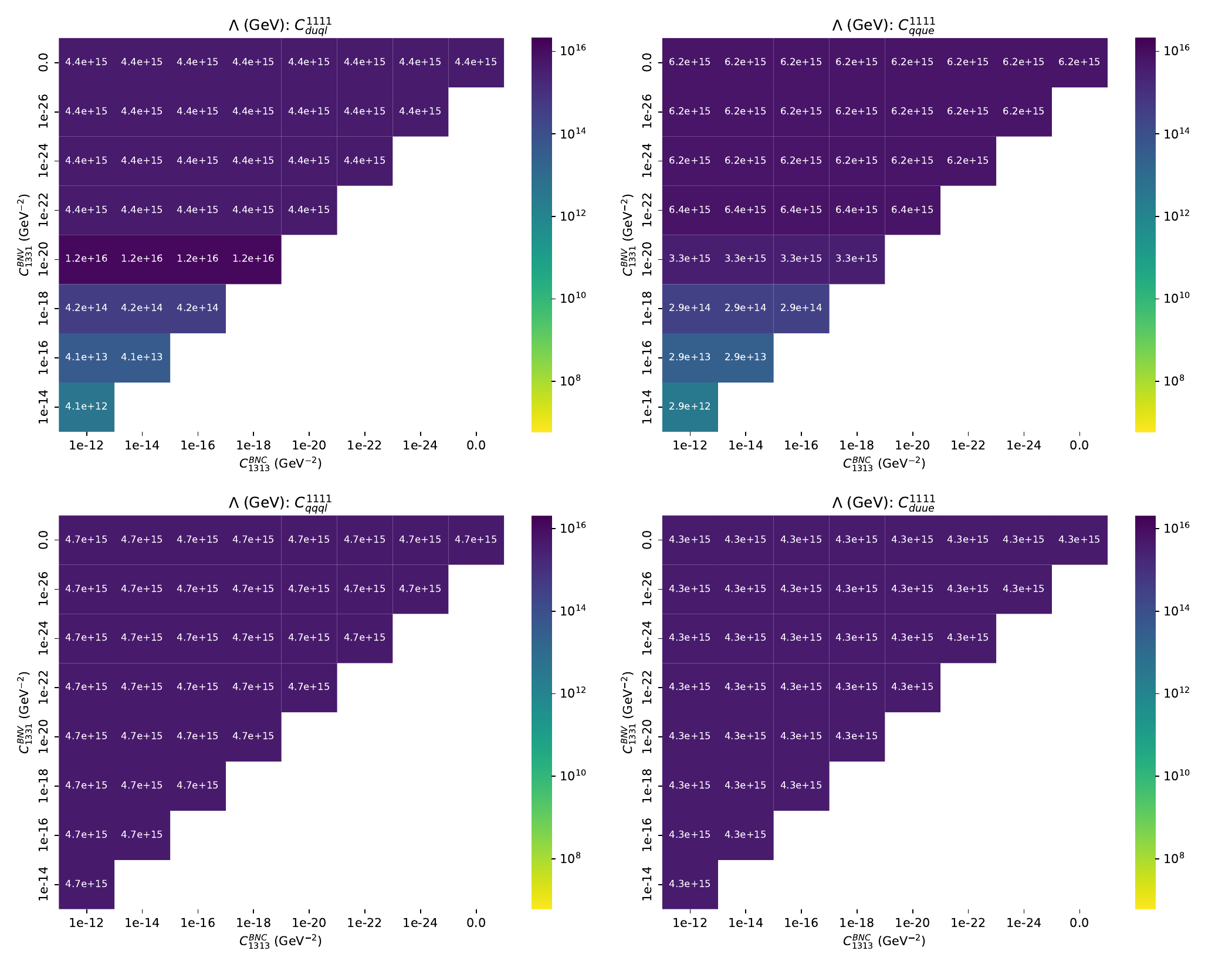}
        \caption{without BNC-BNV mixing terms}
\label{fig:without_2}
    \end{subfigure}
    \begin{subfigure}{1\textwidth}
\centering
        \includegraphics[width=16cm,height=10.5cm]{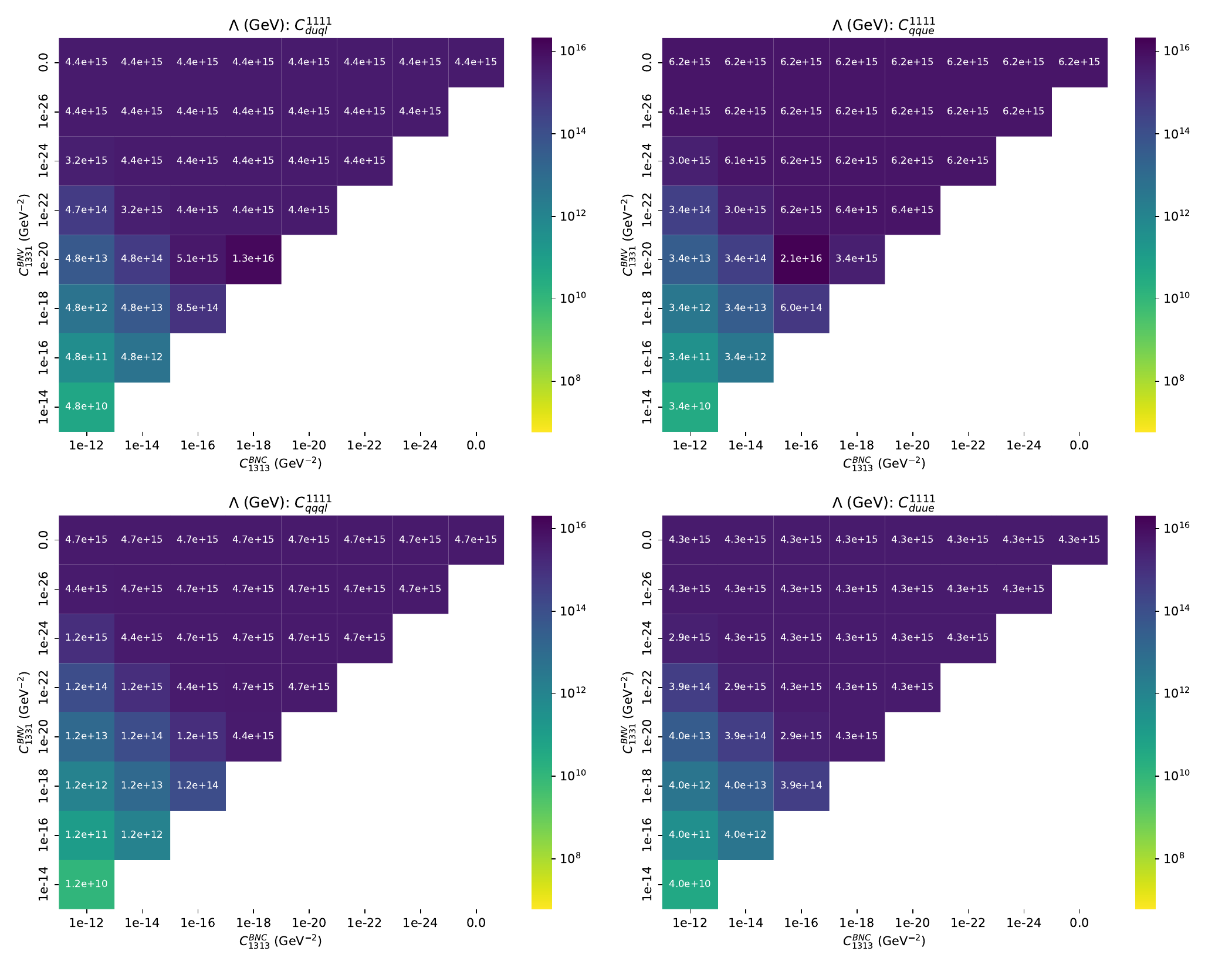}
        \caption{with BNC-BNV mixing terms}
\label{fig:with_2}
    \end{subfigure}
    \caption{The plot shows the dependence of the proton decay operator scale on the BNC-BNV parameter space for case 2. The x-axis represents the BNC operator Wilson coefficient ($C^{BNC}_{1313}$), while, the y-axis represent the BNV Wilson coefficient ($C^{BNV}_{1331}$).}

\label{fig:correlation_2}
\end{figure}

The corresponding UV scale values $\Lambda / \sqrt{c}$ are presented in Fig.~\ref{fig:Lambda_comparison_case2}, comparing results without RG running (in dotted line), with RG evolution in the absence of BNC-BNV mixing (in blue), and with full BNC-BNV mixing (in orange), for different initial scale values of each BNV Wilson coefficient and the numerical limits extracted for them case are listed in Table.~\ref{tab:limits_case_2}. The dependence of the proton decay scale on the BNC-BNV mixing terms is summarised in Fig.~\ref{fig:correlation_2}. The plot Fig.\ref{fig:without_2} shows the proton decay scale obtained after RG evolution without including the BNC-BNV mixing, while Fig.~\ref{fig:with_2} shows the effect of the mixing.

\begin{table}[ht!]
\centering
\resizebox{\textwidth}{!}{%
\begin{tabular}{l|c|c|c|c|c|c|c|c}
\hline
\hline
BNV Operator & \multicolumn{4}{c|}{$Q^{duql}_{1111}$} & \multicolumn{4}{c}{$Q^{qque}_{1111}$}  \\
\hline
\hline
Set of Initial Scale & Set 1 & Set 2 & Set 3 & Set 4  & Set 1 & Set 2 & Set 3 & Set 4   \\
\hline
$\Lambda^{\mathrm{w/o}}/\sqrt{c}$ [in GeV] & \multicolumn{4}{c|}{$3.9\times10^{15}$} & \multicolumn{4}{c}{$5.5\times10^{15}$}  \\
\hline
\multirow{2}{*}{$\Lambda^{\mathrm{w/SM}}/\sqrt{c}$ [in GeV]} & $4.13$ & $4.15$ & $4.4$ & $4.38$ & $2.94$ & $2.94$ & $6.35$ & $6.21$  \\
& $\times10^{12}$ & $\times10^{14}$ & $\times10^{15}$ & $\times10^{15}$ & $\times10^{12}$ & $\times10^{14}$ & $\times10^{15}$ & $\times10^{15}$  \\
\hline
\multirow{2}{*}{$\Lambda^{\mathrm{w/mix}}/\sqrt{c}$ [in GeV]} & $4.75$ & $4.75$ & $4.72$ & $4.36$ & $3.38$ & $3.38$ & $3.37$ & $6.11$  \\
& $\times10^{10}$ & $\times10^{12}$ & $\times10^{14}$ & $\times10^{15}$ & $\times10^{10}$ & $\times10^{12}$ & $\times10^{14}$ & $\times10^{15}$  \\
\hline
\multirow{2}{*}{$\Lambda^{\mathrm{w/SM}}/\Lambda^{\mathrm{w/o}}$} & $1.1$ & \multirow{2}{*}{$0.11$} & \multirow{2}{*}{$1.1$} & \multirow{2}{*}{$1.1$} & $5.3$ & $5.3$ & \multirow{2}{*}{$1.2$} & \multirow{2}{*}{$1.1$}  \\
& $\times10^{-3}$ &  &  &  & $\times10^{-4}$ & $\times10^{-2}$ & &  \\
\hline
\multirow{2}{*}{$\Lambda^{\mathrm{w/mix}}/\Lambda^{\mathrm{w/o}}$}  & $1.2$ & $1.2$ & \multirow{2}{*}{$0.12$} & \multirow{2}{*}{$1.1$} & $6.1$ & $6.1$ & $6.1$ & \multirow{2}{*}{$1.1$}   \\
& $\times10^{-5}$ & $\times10^{-3}$ &  &  &   $\times10^{-6}$ & $\times10^{-4}$ & $\times10^{-2}$ &   \\
\hline
\hline
BNV Operator &  \multicolumn{4}{c|}{$Q^{qqql}_{1111}$} & \multicolumn{4}{c}{$Q^{duue}_{1111}$}   \\
\hline
\hline
Set of Initial Scale &  Set 1 & Set 2 & Set 3 & Set 4  & Set 1 & Set 2 & Set 3 & Set 4  \\
\hline
$\Lambda^{\mathrm{w/o}}/\sqrt{c}$ [in GeV]  & \multicolumn{4}{c|}{$3.9\times10^{15}$} & \multicolumn{4}{c}{$3.9\times10^{15}$} \\
\hline
\multirow{2}{*}{$\Lambda^{\mathrm{w/SM}}/\sqrt{c}$ [in GeV]} & $4.74$ & $4.74$ & $4.74$ & $4.74$ & $4.31$ & $4.31$ & $4.31$ & $4.31$  \\
& $\times10^{15}$ & $\times10^{15}$ & $\times10^{15}$ & $\times10^{15}$ & $\times10^{15}$ & $\times10^{15}$ & $\times10^{15}$ & $\times10^{15}$  \\
\hline
\multirow{2}{*}{$\Lambda^{\mathrm{w/mix}}/\sqrt{c}$ [in GeV]}  & $1.22$ & $1.22$ & $1.22$ & $4.42$ & $3.96$ & $3.96$ & $3.94$ & $4.28$  \\
& $\times10^{10}$ & $\times10^{12}$ & $\times10^{14}$ & $\times10^{15}$ & $\times10^{10}$ & $\times10^{12}$ & $\times10^{14}$ & $\times10^{15}$ \\
\hline
$\Lambda^{\mathrm{w/SM}}/\Lambda^{\mathrm{w/o}}$ & $1.2$ & $1.2$ & $1.2$ & $1.2$ & $1.1$ & $1.1$ & $1.1$ & $1.1$  \\
\hline
\multirow{2}{*}{$\Lambda^{\mathrm{w/mix}}/\Lambda^{\mathrm{w/o}}$}  & $3.1$ & $3.1$ & $3.1$ & \multirow{2}{*}{$1.1$} & $1$ & $1$ & \multirow{2}{*}{$0.1$} & \multirow{2}{*}{$1.1$} \\
&  $\times10^{-6}$ & $\times10^{-4}$ & $\times10^{-2}$ &  &   $\times10^{-5}$ & $\times10^{-3}$ &  &     \\
\hline
\hline
\end{tabular}
}
\caption{Values of the UV scale $\Lambda/\sqrt{c}$ obtained under 4 sets of initial conditions, for case 2, as in Table.~\ref{tab:limits_case_1}, where the running is performed as a function of the energy scale $\mu$ between $10^3$~GeV and $10^{6}$~GeV.}
\label{tab:limits_case_2}
\end{table}

\subsection{Case 3: BNC operator scale at \(10^9\) GeV}\label{subsec:BF-run-109}
In the third benchmark case, we consider even stringent limit on the baryon number conserving (BNC) operators, which is taken to be lower than those adopted in Sec.~(\ref{subsec:BF-run-104}) and Sec.~(\ref{subsec:BF-run-106}). The running of the Wilson coefficients is performed from $10^3$~GeV up to $10^{9}~\text{GeV}$. The corresponding initial scale values for the BNV and BNC operator coefficients are summarised in Table.~\ref{tab:initial_conditions}.

\begin{figure}[htbp!]
     \centering
\hspace*{-0.5cm}\includegraphics[scale=0.55]{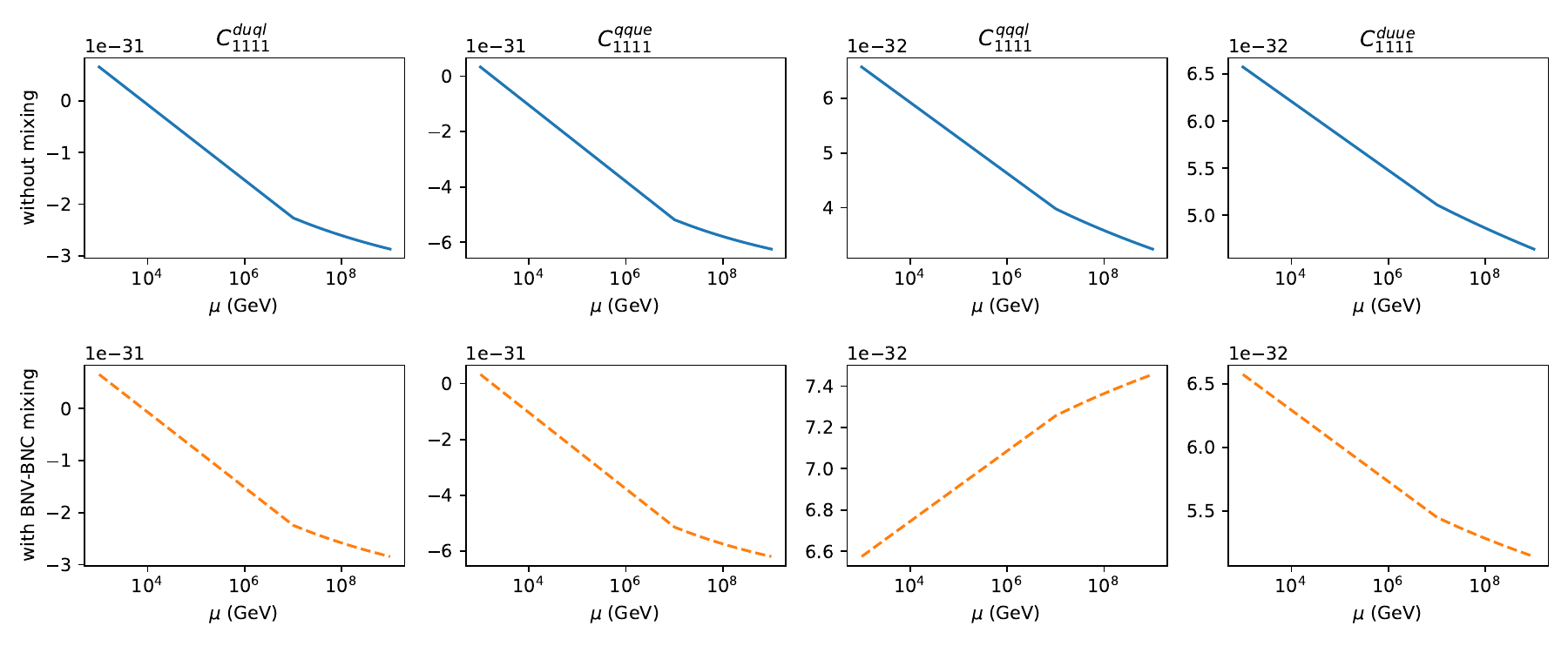}
    \caption{Evolution of the proton decay Wilson coefficients, where the running is performed as a function of the energy scale $\mu$ between $10^3$~GeV and $10^{9}$~GeV. This running corresponds to the case with an initial value of $4\times10^{-20}~\text{GeV}^{-2}$ (Set~1) for the Wilson coefficients $C^{duql}_{1331}$, $C^{qque}_{1331}$, $C^{qqql}_{1331}$, and $C^{duue}_{1331}$.}
    \label{fig:Case3_running}
\end{figure}

Fig.~\ref{fig:Case3_running} shows the evolution of the four proton decay Wilson coefficients $C^{duql}_{1111}$, $C^{qque}_{1111}$, $C^{qqql}_{1111}$, and $C^{duue}_{1111}$ over the energy range. The comparison shows the running without BNC-BNV mixing (blue) and the evolution including full BNC-BNV operator mixing (dotted orange). For the benchmark initial condition with $C^{duql}_{1331} = C^{qque}_{1331} = C^{qqql}_{1331} = C^{duue}_{1331} = 4\times10^{-20}~\text{GeV}^{-2}$ (Set 1), the deviation from the without mixing trajectory is visibly enhanced compared to other cases.

\begin{figure}[ht!]
    \centering
\hspace*{-1.5cm}\includegraphics[scale=0.7]{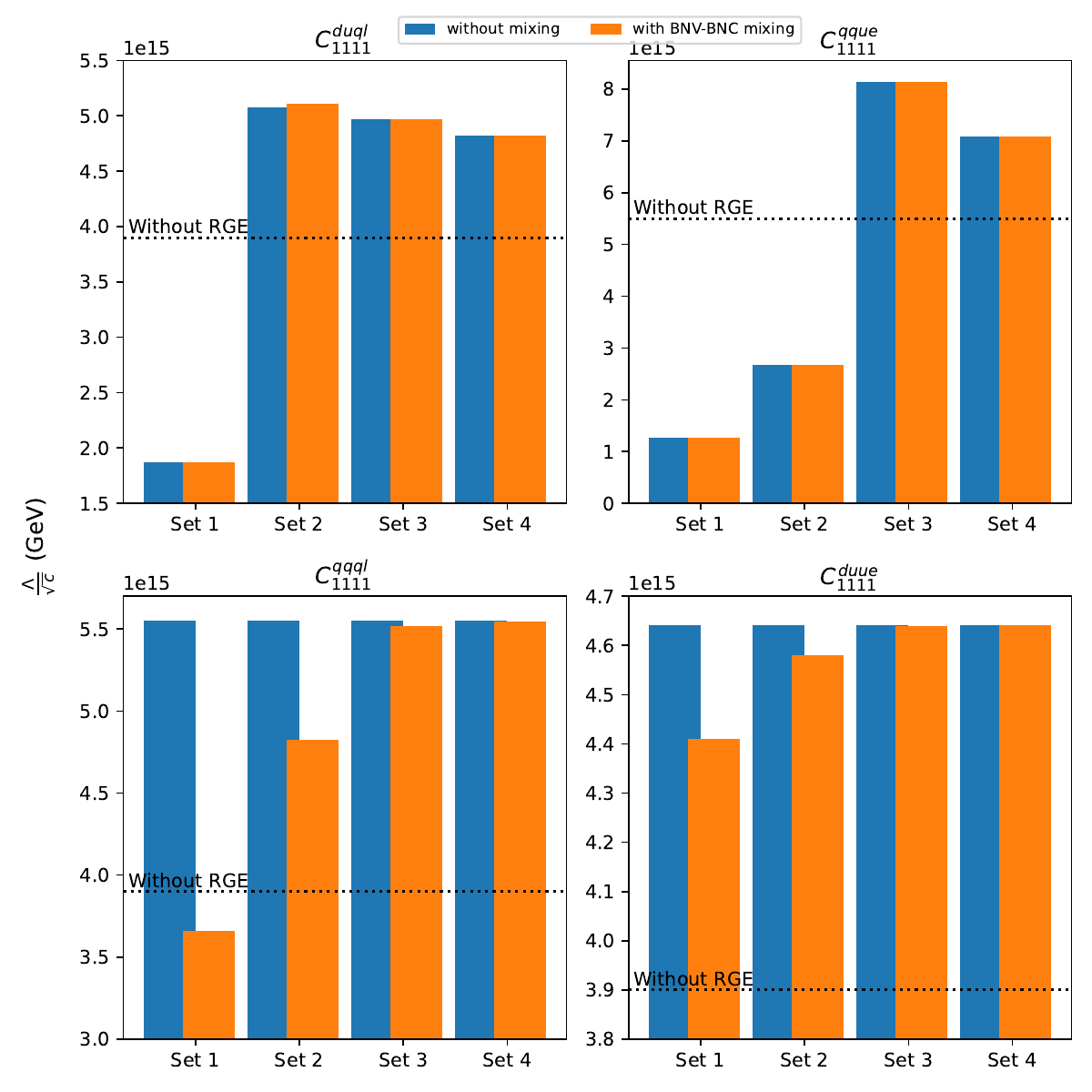}
    \caption{Comparison of the UV scale $\Lambda / \sqrt{c}$ as in Fig.~\ref{fig:Lambda_comparison_case1} and Fig.~\ref{fig:Lambda_comparison_case2}, where the running is performed as a function of the energy scale $\mu$ between $10^3$~GeV and $10^{9}$~GeV, for case 3.}
    \label{fig:Lambda_comparison_case3}
\end{figure}

\begin{figure}[htbp!]
\centering
    \begin{subfigure}{1\textwidth}
\centering
        \includegraphics[width=16cm,height=10.5cm]{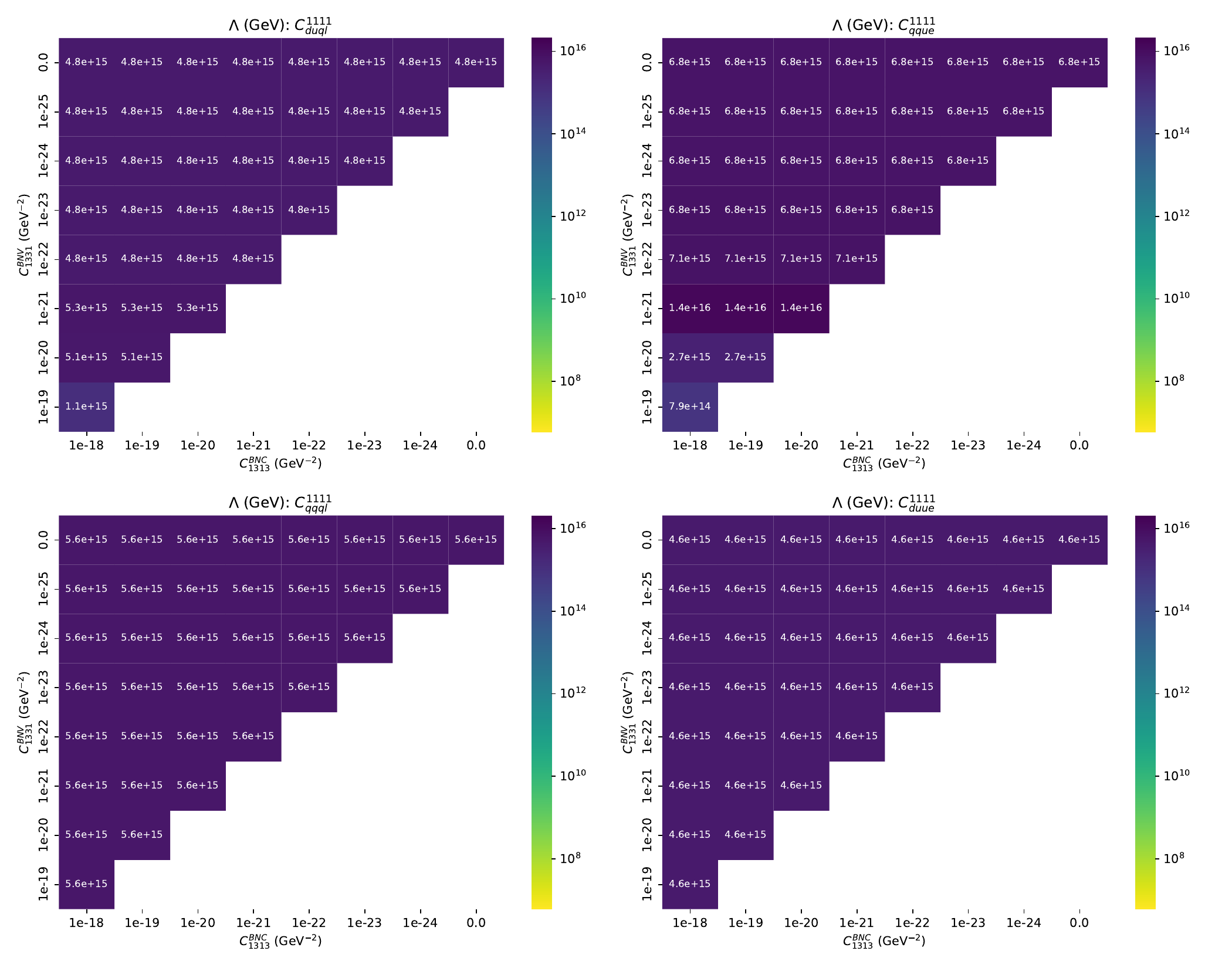}
        \caption{without BNC-BNV mixing terms}
\label{fig:without_3}
    \end{subfigure}
    \begin{subfigure}{1\textwidth}
\centering
        \includegraphics[width=16cm,height=10.5cm]{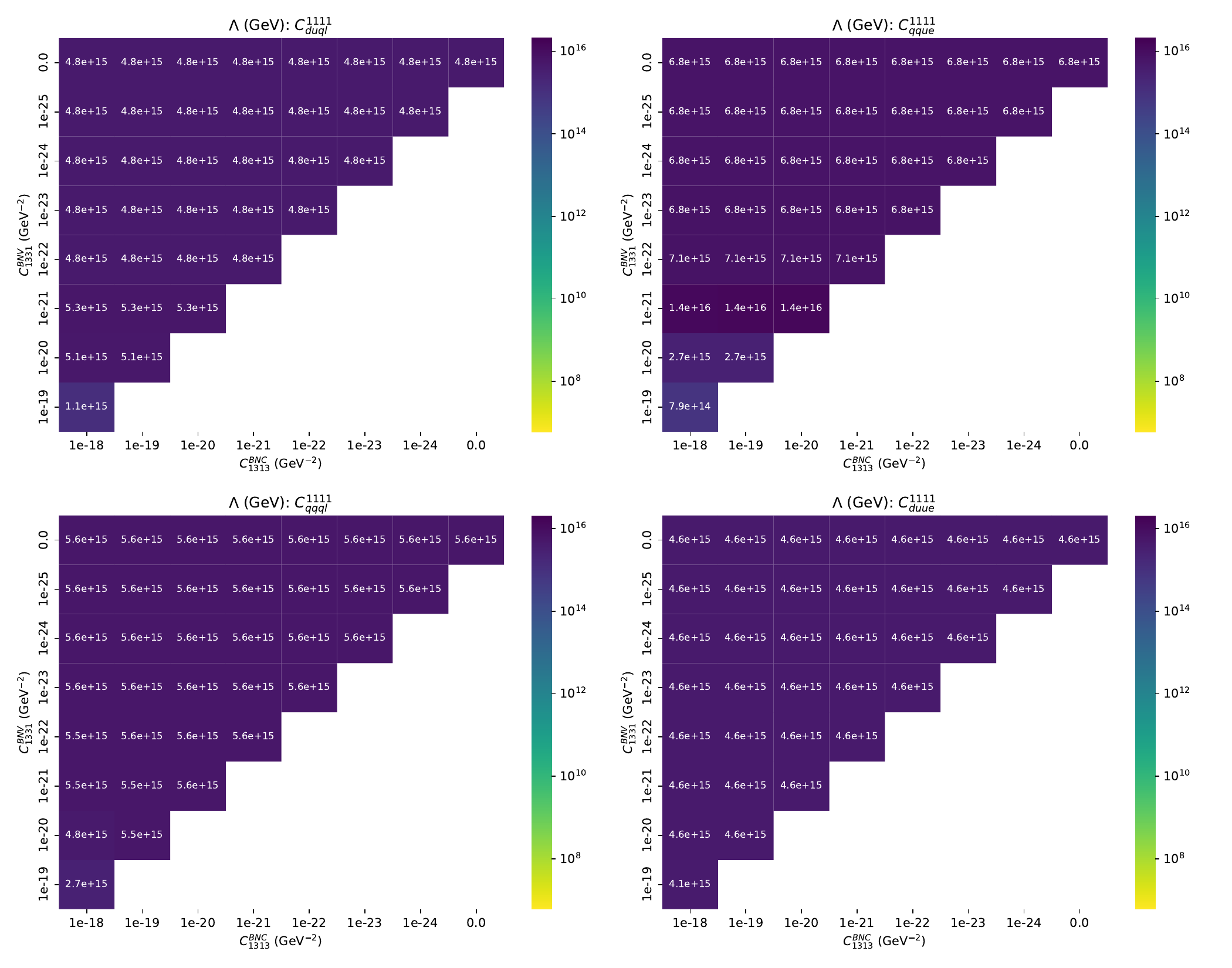}
        \caption{with BNC-BNV mixing terms}
\label{fig:with_3}
    \end{subfigure}
    \caption{The plot shows the dependence of the proton decay operator scale on the BNC-BNV parameter space for case 3. The x-axis represents the BNC operator Wilson coefficient ($C^{BNC}_{1313}$), while, the y-axis represent the BNV Wilson coefficient ($C^{BNV}_{1331}$).}
\label{fig:correlation_3}
\end{figure}

The values of the UV scale $\Lambda / \sqrt{c}$ obtained under three conditions: without RG evolution (in dotted line), with running in scenario without BNC-BNV mixing (in blue), and with full BNC-BNV mixing (in orange) for different initial scale values of each BNV Wilson coefficient are displayed in Fig.~\ref{fig:Lambda_comparison_case3}. The numerical results corresponding to this case are summarised in Table.~\ref{tab:limits_case_3}. The dependence of the proton decay scale on the BNC-BNV mixing terms is summarised in Fig.~\ref{fig:correlation_3}. The plot Fig.\ref{fig:without_3} shows the proton decay scale obtained after RG evolution without including the BNC-BNV mixing, while Fig.~\ref{fig:with_3} shows the effect of the mixing.

\begin{table}[ht!]
\centering
\resizebox{\textwidth}{!}{%
\begin{tabular}{l|c|c|c|c|c|c|c|c}
\hline
\hline
BNV Operator & \multicolumn{4}{c|}{$Q^{duql}_{1111}$} & \multicolumn{4}{c}{$Q^{qque}_{1111}$}  \\
\hline
\hline
Set of Initial Scale & Set 1 & Set 2 & Set 3 & Set 4  & Set 1 & Set 2 & Set 3 & Set 4   \\
\hline
$\Lambda^{\mathrm{w/o}}/\sqrt{c}$ [in GeV] & \multicolumn{4}{c|}{$3.9\times10^{15}$} & \multicolumn{4}{c}{$5.5\times10^{15}$}  \\
\hline
\multirow{2}{*}{$\Lambda^{\mathrm{w/SM}}/\sqrt{c}$ [in GeV]} & $1.87$ & $5.08$ & $4.97$ & $4.82$ & $1.27$ & $2.67$ & $8.15$ & $7.09$  \\
& $\times10^{15}$ & $\times10^{15}$ & $\times10^{15}$ & $\times10^{15}$ & $\times10^{15}$ & $\times10^{15}$ & $\times10^{15}$ & $\times10^{15}$  \\
\hline
\multirow{2}{*}{$\Lambda^{\mathrm{w/mix}}/\sqrt{c}$ [in GeV]} & $1.87$ & $5.11$ & $4.97$ & $4.82$ & $1.27$ & $2.68$ & $8.13$ & $7.08$  \\
& $\times10^{15}$ & $\times10^{15}$ & $\times10^{15}$ & $\times10^{15}$ & $\times10^{15}$ & $\times10^{15}$ & $\times10^{15}$ & $\times10^{15}$  \\
\hline
$\Lambda^{\mathrm{w/SM}}/\Lambda^{\mathrm{w/o}}$ & $0.48$ & $1.3$ & $1.3$ & $1.2$ & $0.23$ & $0.49$ & $1.5$ & $1.3$  \\
\hline
$\Lambda^{\mathrm{w/mix}}/\Lambda^{\mathrm{w/o}}$  & $0.48$ & $1.3$ & $1.3$ & $1.2$ & $0.23$ & $0.49$ & $1.5$ & $1.3$   \\
\hline
\hline
BNV Operator &  \multicolumn{4}{c|}{$Q^{qqql}_{1111}$} & \multicolumn{4}{c}{$Q^{duue}_{1111}$}   \\
\hline
\hline
Set of Initial Scale &  Set 1 & Set 2 & Set 3 & Set 4  & Set 1 & Set 2 & Set 3 & Set 4  \\
\hline
$\Lambda^{\mathrm{w/o}}/\sqrt{c}$ [in GeV]  & \multicolumn{4}{c|}{$3.9\times10^{15}$} & \multicolumn{4}{c}{$3.9\times10^{15}$} \\
\hline
\multirow{2}{*}{$\Lambda^{\mathrm{w/SM}}/\sqrt{c}$ [in GeV]} & $5.55$ & $5.55$ & $5.55$ & $5.55$ & $4.64$ & $4.64$ & $4.64$ & $4.64$  \\
& $\times10^{15}$ & $\times10^{15}$ & $\times10^{15}$ & $\times10^{15}$ & $\times10^{15}$ & $\times10^{15}$ & $\times10^{15}$ & $\times10^{15}$  \\
\hline
\multirow{2}{*}{$\Lambda^{\mathrm{w/mix}}/\sqrt{c}$ [in GeV]}  & $3.66$ & $4.82$ & $5.52$ & $5.54$ & $4.41$ & $4.58$ & $4.64$ & $4.64$  \\
& $\times10^{15}$ & $\times10^{15}$ & $\times10^{15}$ & $\times10^{15}$ & $\times10^{15}$ & $\times10^{15}$ & $\times10^{15}$ & $\times10^{15}$ \\
\hline
$\Lambda^{\mathrm{w/SM}}/\Lambda^{\mathrm{w/o}}$ & $1.4$ & $1.4$ & $1.4$ & $1.4$ & $1.2$ & $1.2$ & $1.2$ & $1.2$  \\
\hline
$\Lambda^{\mathrm{w/mix}}/\Lambda^{\mathrm{w/o}}$  & $0.94$ & $1.2$ & $1.4$ & $1.4$ & $1.1$ & $1.2$ & $1.2$ & $1.2$   \\
\hline
\hline
\end{tabular}
}
\caption{Values of the UV scale $\Lambda/\sqrt{c}$ obtained under the 4 initial conditions, for case 3, as in Table.~\ref{tab:limits_case_1}, where the running is performed as a function of the energy scale $\mu$ between $10^3$~GeV and $10^{9}$~GeV.}
\label{tab:limits_case_3}
\end{table}

\section{Introduction to Python Package}\label{sec:python}
The Python package~\cite{RepoName} is designed to solve systems of first-order coupled differential equations for the Wilson coefficients and other Standard Model (SM) parameters. It allows users to specify which operators will run, along with any additional variables required to evaluate the corresponding beta functions.

The beta functions for the running variables are read from an external text file provided as input to the solver, making the framework flexible and easily customisable. The solver automatically handles redundant Wilson coefficients. For instance, $C_{3131}^{qu1}$ and $C_{1313}^{qu1}$ are complex conjugates of each other, and the solver ensures that this condition is maintained at all times.
The solver is implemented as a Python class, allowing users to create multiple instances with different initial conditions and run them independently. Once a solver instance is created, the user must:
\begin{enumerate}
    \item Provide the list of variables to be evolved,
    \item Specify the initial conditions for those variables,
    \item Define the initial and final energy scales.
\end{enumerate}

This setup is achieved using the various methods (member functions) defined in the class. If the initial conditions or the energy scale are not specified, the solver uses its default values.
Note that since Python arrays are zero-indexed, the indices $(prst)$ are zero-based; for example, $C_{3131}^{qu1} \rightarrow \texttt{Cqu1[2,0,2,0]}$.

After initialization, the user can call another method to load the beta functions from the input text file. This method constructs a dictionary of beta functions for all variables to be evolved by substituting the current values of all other parameters.

The system of coupled differential equations is then solved using the \texttt{scipy} library's built-in numerical solvers. The resulting solution is stored as a dense output variable within the class, enabling continuous evaluation over the entire energy range.

To extract results at specific energy scales, the user can call a separate method that evaluates this dense output at the desired points.

\section{Summary}\label{sec:Summary}
The proton decay scale in simple grand unified models is constrained to be around $\sim 10^{15}~\text{GeV}$ by current nucleon decay lifetime experiments. However, such estimates do not account for the RG running of the corresponding Wilson coefficients. Usually, in literature, effect of RG running involving corrections from SM at next-to-leading order is discussed, assuming no new-physics exists between the electro-weak sector and the proton decay scale. Nevertheless, this notion of existence of a large desert between the electroweak and BNV scales is increasingly disfavoured.

The RG evolution should, therefore, be supplemented by all new-physics that may arise at intermediate energy thresholds between these scales. In a model independent SMEFT framework, dimension-6 BNC operators at below these intermediate thresholds contribute to the RG running of the proton decay Wilson coefficients. The dominant 1-loop contribution to the BNC-BNV mixing arises from operators with top quark loops. Since collider experiments do not constrain flavour violating BNC operators involving top quarks beyond $\sim 10^{4}~\text{GeV}$, this motivates a detailed study of the RG evolution of the Wilson coefficients for proton decay operators, incorporating possible mixing. This is further supported by the fact that BNV operators containing third-generation quarks are themselves not constrained, allowing their scales to be set phenomenologically. Consequently, although the mixing between two dimension-6 operators is naively expected to be subleading compared to that of dimension-seven operators, such mixing can nonetheless induce significant modifications to the corresponding beta functions.

We show that BNC-BNV operator mixing, particularly involving third-generation quarks through flavour violating operators, can significantly enhance the RG running effects and thereby lower the effective new-physics scale for generating for proton decay ($\sim 10^7$ GeV). Correlation and the dependence of the proton decay scale on BNC-BNV Wilson coefficients is shown in Fig.\ref{fig:correlation_1}, Fig.\ref{fig:correlation_2} and Fig.\ref{fig:correlation_3}. This highlights the importance of incorporating flavor violation structures and higher-dimensional SMEFT contributions in nucleon decay analyses. To facilitate further exploration, we provide a Python implementation that performs the RG evolution of nucleon decay Wilson coefficients up to arbitrary SMEFT scales, allowing for generic baryon number conserving extensions. The evolved coefficients are then compared with experimental bounds on nucleon decay lifetimes to constrain the scale and structure of baryon number violating interactions.

\FloatBarrier
\section*{Acknowledgments}
M.T.A. acknowledges the financial support of DST through the INSPIRE Faculty grant DST/INSPIRE/04/2019/002507. 

\section*{Appendix}
\appendix
\section{Representative 1-loop mixing calculation}
\label{app:loop-calc}
In this Appendix, we present a representative 1-loop calculation illustrating the extraction of the ultraviolet-divergent part that determines the BNC-BNV mixing coefficient \(a_{ij}\). The example corresponds to the diagram topology (d) shown in Fig.~\ref{fig:BNV-BNC-loops}, involving the loop formed between the operators \(Q^{duql}_{1331}\) and \(Q^{qu1}_{1313}\) for the BNV operator \(Q^{duql}_{1111}\). In this configuration, a BNV operator is inserted on one fermion line and a BNC operator on the other, with the two connected through a closed top quark loop. The same procedure applies to other operator pairs with appropriate modifications to their Dirac and colour structures, and the complete set of contributions is summarised in Table.~\ref{tab:mixing_terms}.

\begin{figure}[h!]
    \centering
    \includegraphics[scale=0.17]{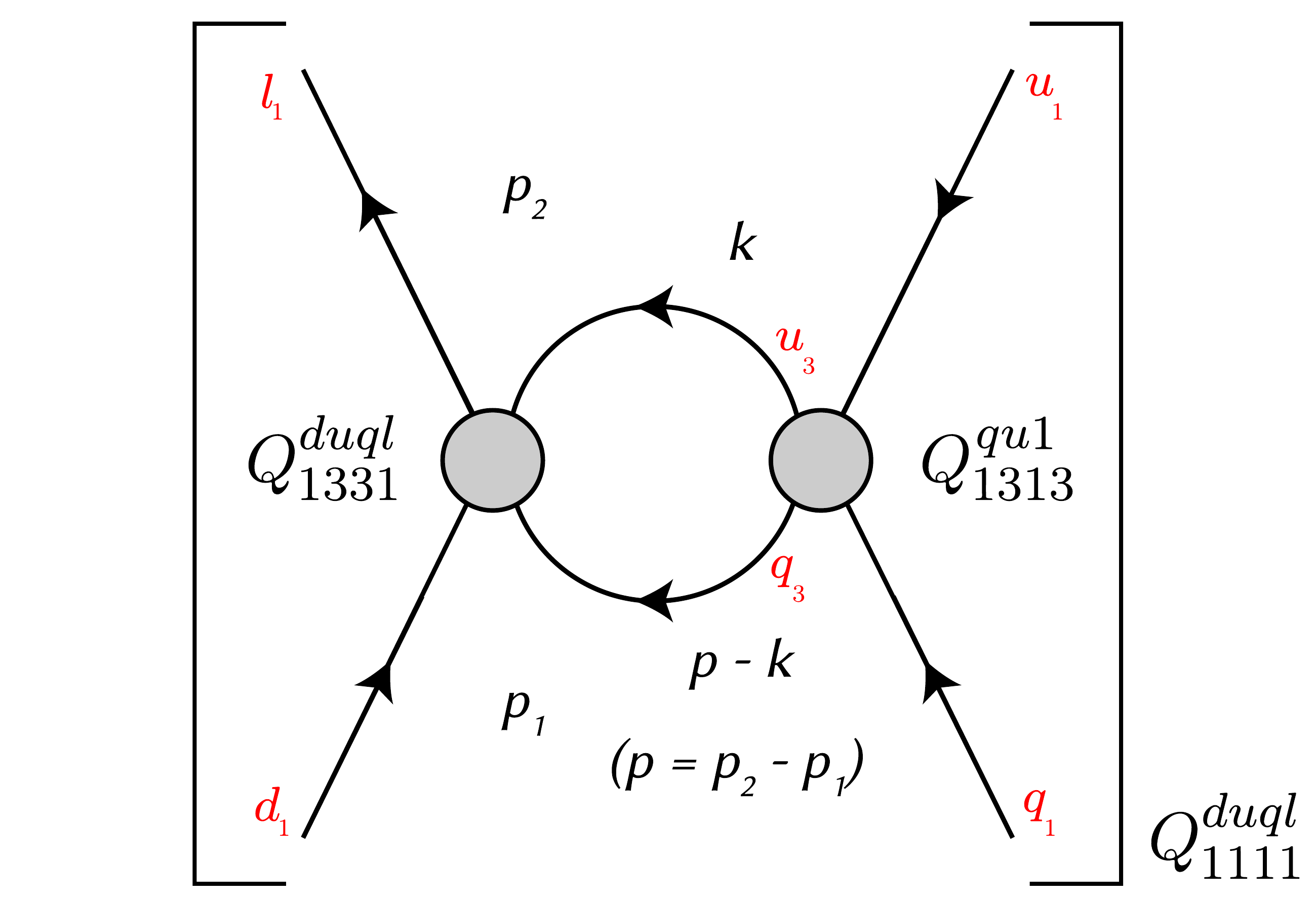}
    \caption{One-loop diagram illustrating the mixing between the operators \(Q^{duql}_{1331}\) and \(Q^{qu1}_{1313}\) through a closed top quark loop for the BNV operator \(Q^{duql}_{1111}\).}
    \label{fig:loop_diagram}
\end{figure}

The corresponding 1-loop diagram is depicted in Fig.~\ref{fig:loop_diagram}, where the flow of external and internal momenta is explicitly indicated. The external momenta of the incoming and outgoing quark fields are denoted by \(p_{1}\) and \(p_{2}\), respectively, while the loop momentum is labelled \(k\). The momentum carried by the propagator connecting the two operator insertions is therefore \(p - k\), with \(p = p_{2} - p_{1}\). The fermion flow follows the convention shown in the figure.
The coloured labels (in red) identify the quark flavours and their corresponding generation indices appearing in the operators. Specifically, the diagram involves a top quark loop connecting the \(u_{3}\) and \(q_{3}\) fields, consistent with the flavour structure of the chosen operator combination \(Q^{duql}_{1331}\) and \(Q^{qu1}_{1313}\).

The associated 1-loop Feynman integral is then expressed as
\begin{equation}
    I = -C^{qu1}_{1313} \, C^{duql}_{1331}\int \! \frac{d^{d}k}{(2\pi)^{d}} \; \frac{\mathrm{Tr}\!\left[\gamma_{\mu}\gamma^{\mu}(\slashed{k}+m_t)\,(\slashed{p}-\slashed{k}+m_t)\right]} {(k^{2}-m_t^{2})\,[(p-k)^{2}-m_t^{2}]},
\end{equation}
where \(m_t\) denotes the mass of the top quark.

After evaluating the Feynman integral in dimensional regularisation with \(d = 4 - 2\epsilon\), and neglecting the external momentum \(p\) compared to the heavy top quark mass \(m_t\), the divergent part of the loop integral reduces to
\begin{equation}
    I = -\frac{12\,m_t^{2}\,\Delta_{\epsilon}}{16\pi^{2}}C^{qu1}_{1313} \, C^{duql}_{1331} + \mathcal{O}(\epsilon),
\end{equation}
where,
\begin{equation}
    \Delta_{\epsilon} = \left( \frac{1}{\epsilon} - \gamma_{E} + \ln 4\pi \right),
\end{equation}
with $\gamma_E$ being the Euler-Mascheroni constant ($\gamma_E$ = 0.5772156649).
The ultraviolet pole \(1/\epsilon\) captures the divergent contribution relevant for operator mixing. Finite terms are scheme dependent and omitted here, since only the divergent part determines the renormalisation of the BNV operator through the mixing coefficient \(a_{ij}\). Hence, 
\begin{equation}
    a^{duql}_{ij} \, C^{qu1}_{i,1313} \, C^{duql}_{j,1331} = -\frac{12\,m_t^2}{16 \pi^2} C^{qu1}_{i,1313} \, C^{duql}_{j,1331}.
\end{equation}

\section{Index of Notations}
\label{app:index}
\renewcommand{\arraystretch}{1.3}
\begin{longtable}{p{3cm} p{12cm}}
\textbf{Symbol} & \textbf{Description} \\ \hline

$\Lambda$ & New-physics scale suppressing higher-dimensional operators in SMEFT. \\

$L_{\text{SMEFT}}$ & Full SMEFT Lagrangian including dimension-d (for $d>4$) extensions: $L_{\text{SMEFT}} = L_{\text{SM}} + \sum_{d>4}\frac{1}{\Lambda^{d-4}}L^{(d)}$. \\

$L^{(6)}$ & dimension-6 part of the SMEFT Lagrangian, containing all operators suppressed by $\Lambda^2$. \\

$p,r,s,t,v,w$ & Flavour (generation) indices, taking values $1,2,3$ for the three SM families. \\

$\ell, q, u, d, e$ & Lepton doublet, quark doublet, right-handed up-type quark, right-handed down-type quark, and charged lepton fields respectively. \\

$Q_{i,prst}$ & Gauge-invariant dimension-6 operator in the Warsaw basis; indices $p,r,s,t$ denote fermion generations. \\

$C_{i,prst}$ & Dimensionful Wilson coefficient of operator $Q_{i,prst}$, related to the dimensionless coefficient $c_{i,prst}$ by $C_{i,prst} = c_{i,prst}/\Lambda^2$. \\

$c_{i,prst}$ & Dimensionless Wilson coefficient corresponding to $Q_{i,prst}$. \\

$\Delta B, \Delta L$ & Change in baryon number and lepton number carried by the operator. \\

$Q^{\text{BNC}}_{i,prst}$ & Baryon Number Conserving (BNC) dimension-6 operators built from SM fields. \\

$Q^{\text{BNV}}_{i,prst}$ & Baryon Number Violating (BNV) operators changing $\Delta B = \Delta L = 1$. \\

$C^{\text{BNC}}_{i,prst}$ & Wilson coefficients of the BNC operators. \\

$C^{\text{BNV}}_{i,prst}$ & Wilson coefficients of the BNV operators. \\

$\gamma_{kl}$ & 1-loop anomalous dimension matrix dictating RG mixing between operators $Q_k$ and $Q_l$. \\

$a_{ij}$ & Loop-induced BNC-BNV mixing coefficients appearing in $\sum_{i,j} a_{ij} C^{\text{BNC}}_i C^{\text{BNV}}_j$. \\

$\Delta_{\epsilon}$ & Divergent factor in dimensional regularisation:
$\displaystyle \Delta_\epsilon = \left( \frac{1}{\epsilon} - \gamma_E + \ln 4\pi \right)$,  
with $\gamma_E$ being the Euler-Mascheroni constant. \\

$\epsilon$ & Regulator parameter in dimensional regularisation, with $d = 4 - 2\epsilon$. The pole $1/\epsilon$ captures UV divergences. \\

$\gamma_E$ & Euler-Mascheroni constant ($\gamma_E$ = 0.5772156649) appearing in the dimensional-regularisation scheme. \\

$g_1, g_2, g_3$ & Gauge couplings corresponding to $U(1)_Y$, $SU(2)_L$, and $SU(3)_c$ respectively. \\

$Y_u, Y_d, Y_e$ & Yukawa coupling matrices for up-type quarks, down-type quarks, and charged leptons. \\

$\epsilon_{\alpha\beta\gamma}$ & Antisymmetric tensor in colour space used to contract $SU(3)_c$ indices in baryon number violating operators. \\

$\epsilon_{ij}$ & Antisymmetric tensor in $SU(2)_L$ space used to contract weak indices in four-fermion operators. \\

$C$ & Charge conjugation matrix acting on spinors. \\

$Q^{duq\ell}_{prst}$ & BNV operator $\epsilon_{\alpha\beta\gamma}\epsilon_{ij}(d^\alpha_p C u^\beta_r)(q^{i\gamma}_s C \ell^j_t)$. \\

$Q^{qque}_{prst}$ & BNV operator $\epsilon_{\alpha\beta\gamma}\epsilon_{ij}(q^{i\alpha}_p C q^{j\beta}_r)(u^\gamma_s C e_t)$. \\

$Q^{qqq\ell}_{prst}$ & BNV operator $\epsilon_{\alpha\beta\gamma}\epsilon_{il}\epsilon_{jk}(q^{i\alpha}_p C q^{j\beta}_r)(q^{k\gamma}_s C \ell^l_t)$. \\

$Q^{duue}_{prst}$ & BNV operator $\epsilon_{\alpha\beta\gamma}(d^\alpha_p C u^\beta_r)(u^\gamma_s C e_t)$. \\

$\dot{C}_k$ & Renormalization group (RG) derivative: $\dot{C}_k = 16\pi^2 \mu \frac{dC_k}{d\mu} = \sum_l \gamma_{kl} C_l$. \\

$\mu$ & Renormalization scale used in RG evolution. \\

$m_t$ & Mass of the top quark, relevant for dominant top-loop contributions. \\

Set of Initial values & Refers to different benchmark configurations of initial Wilson coefficients at the reference scale, used to study RG evolution scenarios. \\

$\Lambda / \sqrt{c}$ & BNV effective ultraviolet (UV) scale associated with a given Wilson coefficient $C_{i,prst} = c_{i,prst}/\Lambda^2$. \\

$\Lambda^{\mathrm{w/o}}/\sqrt{c}$ & BNV UV scale obtained \emph{without} RG evolution; tree-level estimate assuming static Wilson coefficients. \\

$\Lambda^{\mathrm{SM}}/\sqrt{c}$ & BNV UV scale including RG running without BNC-BNV mixing contributions. \\

$\Lambda^{\mathrm{mix}}/\sqrt{c}$ & BNV UV scale including both SM and BNC-BNV mixing effects during RG running. \\

$\Lambda^{\mathrm{SM}} / \Lambda^{\mathrm{w/o}}$ & Ratio comparing the effect of pure SM running to the static (no-RG evolution) case. Quantifies the enhancement or suppression due to without-mixing evolution. \\

$\Lambda^{\mathrm{mix}} / \Lambda^{\mathrm{w/o}}$ & Ratio comparing the mixed (SM along with BNC-BNV) evolution to the static case, indicating the total effect of BNC-BNV operator mixing. \\
\end{longtable}

\bibliographystyle{JHEPCust}
\bibliography{vecB}

\end{document}